**RESEARCH** **Open Access**

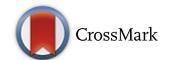

# Lessons learned from applying social network analysis on an industrial Free/Libre/Open Source Software ecosystem

Jose Teixeira[1*], Gregorio Robles[2,3] and Jesús M. González-Barahona[2,3]

**Abstract**

Many software projects are no longer done in-house by a single organization. Instead, we are in a new age where software is developed by a networked community of individuals and organizations, which base their relations to each other on mutual interest. Paradoxically, recent research suggests that software development can actually be jointly-developed by rival firms. For instance, it is known that the mobile-device makers Apple and Samsung kept collaborating in open source projects while running expensive patent wars in the court. Taking a case study approach, we explore how rival firms collaborate in the open source arena by employing a multi-method approach that combines qualitative analysis of archival data (QA) with mining software repositories (MSR) and Social Network Analysis (SNA). While exploring collaborative processes within the OpenStack ecosystem, our research contributes to Software Engineering research by exploring the role of groups, sub-communities and business models within a high-networked open source ecosystem. Surprising results point out that competition for the same revenue model (i.e., operating conflicting business models) does not necessary affect collaboration within the ecosystem. Moreover, while detecting the different sub-communities of the OpenStack community, we found out that the expected social tendency of developers to work with developers from same firm (i.e., homophily) did not hold within the OpenStack ecosystem. Furthermore, while addressing a novel, complex and unexplored open source case, this research also contributes to the management literature in coopetition strategy and high-tech entrepreneurship with a rich description on how heterogeneous actors within a high-networked ecosystem (involving individuals, startups, established firms and public organizations) joint-develop a complex infrastructure for big-data in the open source arena.

## 1 Introduction

Software is often no longer developed in-house, but in an open ecosystem, where employees of a company co-operate with "distributed collaborators" from other companies. If the software ecosystem is sufficiently open, the co-operation will include as well a community of volunteers, developers employed by other partner companies, universities, and even competitors.

Such an inter-networked approach to software development also applies to Free/Libre/Open Source Software (FLOSS) projects which have a strong industrial component. WebKit is an example of such a situation, where several dozen companies collaborate in the development of a web-layout software used in many web browsers. OpenStack, a cloud infrastructure developed together by over 200 companies and organizations, is another case of such an inter-networked software ecosystem.

In an environment of technological complexity and competitive turbulence, companies collaborate with other companies that compete in the same market [1, 2]. For instance, WebKit has been developed together by Google, Apple and Samsung, which were involved in a fierce competition on the smartphone and smartphone operating system markets. It is now well-known that Apple and

*Correspondence: jose.teixeira@utu.fi
[1]School of Economics (TSE), University of Turku, Turku, Finland
Full list of author information is available at the end of the article

Samsung continued their collaboration in WebKit while running expensive patent-wars in the court worldwide [3].

This hybrid behavior comprising competition and cooperation has been named *coopetition*. A number of management scholars have emphasized the increasing importance of coopetition both for today's networked business and networked scientific investigation [4–7]. Nevertheless, it remains undetermined whether such knowledge can contribute to Software Engineering research, which more recently embraced the ecosystem thinking [8].

In this research, we study coopetition in the open source arena by investigating the OpenStack ecosystem that jointly develops, promotes and exploits a complex cloud computing infrastructure. We take software ecosystems as they were early defined by Jansen et al. as "a set of businesses functioning as a unit and interacting with a shared market for software and services, together with the relationships among them" [9].

Within an exploratory case study [10], we employ a mixed methods approach which bridges qualitative analysis of archival data (QA) with mining software repositories (MSR) and Social Network Analysis (SNA) to assess how competition and collaboration among firms evolve over time. This blended methodological approach allowed us to exploit synergies among methods while minimizing single-source and single-method biases [11–13].

Taking a longitudinal design covering more than 4 years of the OpenStack development, we address a call from Basole [14] for the use of methods which take into consideration the time, pace and sequence in the study of an ecosystem. We are also addressing the novel and more specific term of "open-coopetition", coined by Teixeira and Lin as "a portmanteau of cooperative competition in the open source arena, where R&D is jointly performed by competing firms in a open source way, giving-up authorship-granted intellectual property rights for maximizing both the blueprints transparency and collaborative benefits" [3].

We assess how revenue models affect collaboration in OpenStack: contrary to expectations, firms competing for the same revenue model (i.e., where rivalry is expected) tend to collaborate more than firms which do not compete for the same revenue model; being the exception only those firms that provide public cloud services (in the case of OpenStack, these are HP, Rackspace and Canonical).

Our findings indicate that by combining QA methods with SNA visualizations, we can produce rich longitudinal descriptions and interpretations which enable a better understanding of competitive and collaborative issues in large and complex software ecosystems.

Finally, we also contribute to management and innovation studies with a rich description of how heterogeneous actors within a high-networked ecosystem (involving individuals, startups, established firms and public organizations) joint-develop a complex infrastructure for big-data in the open source arena.

## 2 Coopetition and Free/Libre/Open Source software ecosystems

A number of management scholars [4–7] have emphasized the increasing importance of coopetition both for today's networked business and networked scientific investigation. According to Dagnino and Padula [1], the term "coopetition" was coined by Nadar, former CEO of Novell, and introduced into research by Brandenburger, first in an academic journal [15] and then in a more practice-oriented book [7].

The current coopetition body of knowledge argues that competitors can be involved in both cooperative and competitive relationships with each other simultaneously while benefiting from both in a symbiotic way. Coopetitive relationships are complex and hard to manage, as they consist of two diametrically opposed logics of interaction [16]. According to Bengtsson and Kock [2], firms tend to cooperate more frequently in activities carried out at a greater distance from buyers, and to compete in activities closer to buyers. From a strategic point of view, this means that R&D activities, such as software development, are best suited to be developed in cooperation with a competitor, but when it comes to marketing a new product, competitors choose to distinguish the products from each other. A core driving force behind this behavior is the heterogeneity of resources, as each competitor holds unique resources that are best utilized in combination with other competitors' resources. Other driving forces are shorter product life cycles, convergence of multiple technologies and increasing R&D and capital expenditures [5]; rapidly changing consumer preferences; and the speed and magnitude of technological changes [17]; additionally, firms need to speed-up their innovation efforts [18] and to aim at setting up standards and platforms [19, 20].

Even though the existing literature addressing coopetition in the technological sector is still scarce, there is a growing stream of research addressing coopetition grounded on empirical material from the technological sector. For instance, Osarenkhoe described how Nokia, Ericsson, and Motorola cooperated to improve the Chinese telecom infrastructure while competing on the same market with different mobile devices [21]. On the LCD TV markets, Sony and Samsung cooperated strongly in R&D and manufacturing while commercializing innovative flat screen TVs. Addressing collaboration among high-tech giants, Gnyawali and Park [5] found that both Sony and Samsung were able to reap major benefits from applying coopetitive elements in their strategy. By examining data from Taiwanese firms in the information and communication technology industry, Huang and Yu [22] suggested that coopetition in R&D boosts innovation in a firm. Addressing the manufacturing of telecommunication satellites, one of the most competitive segments of the space aircraft industry, Lopez-Fernandez et al. [23] pinpointed that coopetition is filled with tension due to inherent contradictory and opposing forces. They contributed with a conceptual framework that increases the understanding of tension in coopetition, and key approaches to cope with it. Using empirical data from the semiconductor industry Park et al. [24] examined coopetition and its effects on innovation performance. The authors conclude that competition and cooperation intensities have a non-monotonic positive relationship with firms' coopetition-based innovation performance.

Since coopetition applies to inter-firm relationships, the phenomenon is to be observed in inter-organizational networks, which many scholars more recently refer to by using the ecosystems metaphor [14, 25–27]. Following a trend from industry, there is an emergence of Software Engineering research with an interest in software ecosystems [28–30]. It is recognized that "ecosystem thinking" brought a radical shift in how Software Engineering is being carried out, influencing fundamental aspects such as control, collaboration, business models, and innovation [8, 31, 32].

Knowledge of business ecosystems, or even of natural ecosystems, can also be useful for understanding software ecosystems [28–30]. However, as pointed out by Hanssen and Dybå, the software business has radically different characteristics (e.g., short distance from design to use, the intangibility of software, the high innovation velocity) [30]. Hence, it is not clear to what extent theories drawn from business and natural ecosystems apply in the Software Engineering world.

Even if coopetition is a phenomenon that has an impact on how R&D operations are conducted in an ecosystem setting, there are very few empirical studies addressing how rival software development teams simultaneously collaborate and compete [33]. In Software Engineering, empirical research exploring the inherent notions of competition and collaboration in software development teams is as well very scarce. Such scarcity is a principal *raison d'être* for this research.

There are several known cases of open-coopetition (i.e., coopetition in the open source arena) as captured in Table 1. The phenomenon where competing firms jointly develop open source ecosystems has reached different R&D intensive sectors, following the development of the Internet, cloud computing, mobile-devices and automotive technologies among others.

A key concept which emerged in this research was the sociological concept of homophily - the tendency of individuals to associate and bond with similar others [34–37]. A vast array of network studies pointed out that humans tend to connect with humans sharing similar attributes (e.g., age, gender, class, organizational role or company affiliation) [34, 35]. The same applies to the natural ecosystems where animal species tend to mate with similar ones [38]. Prior research dealing with inter-organizational networks, strongly suggests that homophily drives the formation of corporate strategic alliances [39, 40], the formation of entrepreneurial teams [41] or the selection of human resources [42, 43].

The concept of homophily was already addressed in technology related contexts. The concept was integrated in the 'diffusion of innovation' theory [44], as acknowledged previously by Software Engineering scholars [45, 46], "one of the most distinctive problems in the diffusion of innovations is that the participants are usually quite heterophilous" [44]. Existing studies in Information Systems also suggest that homophily plays a very important role in the assimilation of technology [47]. However, the concept remains largely unexplored in studies taking into account both the social and the technological perspectives.

## 3 Research questions

Our research questions focus on understanding the hybrid behaviors of collaboration and competition within the development of the OpenStack Nova open source project, an example of a high-tech industrial Free/Libre/Open Source software ecosystem involving competing firms that market similar products and services.

**RQ1** – *What is the software development process used to develop the OpenStack high-networked open source cloud computing infrastructure?*

**RQ2** – *Are developers affiliated with different firms collaborating with each other in the project? How does the collaboration evolve over time? How is collaboration affected by exogenous events in the market?*

**RQ3** – *Is there a tendency towards sub-grouping in the project? Are there different sub-communities within the OpenStack ecosystem-community? Which ones? How do developers cluster into different groups? Do developer clusters correspond to firms?*

**RQ4** – *Do firms that compete in the same revenue model collaborate less in the ecosystem?*

## 4 Case Study: OpenStack

OpenStack is an open source software cloud computing platform that is primarily deployed as an "Infrastructure as a Service" (IaaS) solution. It started as a joint project of Rackspace, an established IT web hosting company, and NASA, the well-known USA governmental agency responsible for the civilian space program, aeronautics and aerospace research. Today more than 200 firms and many individual contributors contribute to a set of different open source projects governed by the OpenStack Foundation.

**Table 1** Coopetition in the open source arena

| Project | Project domain | Competing firms collaborating in the project |
|---|---|---|
| WebKit | Web-browsing technologies | Apple, Nokia, Google, Samsung, Intel, RIM among others. |
| Blink | Web-browsing technologies | Google, Opera, Intel, Samsung among others. |
| OpenStack | Cloud computing infrastructure | Rackspace, Canonical, IBM, HP, Vmware, Citrix among others. |
| Cloud Foundry | Platform as a Service (PaaS) | Cisco, Canonical, IBM, EMC, VMware, SAP among others. |
| Xen | Virtualization | Citrix, IBM, Intel, HP, Novell, Red Hat, Oracle among others. |
| Open Handset Alliance | Mobile devices platform | Asus, LG, Samsung, HTC, Acer, Huawei, ZTE among others. |
| Tizen | Operating System | Fujitsu, Huawei, NEC, Casio, Panasonic, Samsung among others. |
| GENIVI Alliance | In-Vehicle Infotainment | Volvo, BMW, Honda, Hyundai, Renault, PSA among others. |

Both hardware and software developers affiliated with companies such as AT&T, AMD, Canonical, Cisco, Dell, EMC, Ericsson, HP, IBM, Intel, NEC, NASA and many others, work together with independent, non-affiliated developers in a scenario of pooled R&D in an open source fashion. We decided to address the OpenStack case due to its perceived novelty, its high inter-networked nature (i.e., an ecosystem involving many firms and individual contributors), its heterogeneity (i.e., an ecosystem involving both startups and high-tech corporate giants), its market-size ($1.7bn by 2016 as claimed by Al Sadowski 451 Research analyst in August of 2014 [48]), its complexity (i.e., involving different programming languages, different operating systems, different hardware configurations) and its size (17,020 community members, 100,000 code reviews and 1,766,546 lines of code as recently reported by Jason Baker from Red Hat in June 2014 [49]).

The OpenStack Nova project, our unit of analysis, is a cloud computing fabric controller, the main part of an IaaS system. It is the largest and the most "core" project governed by the OpenStack Foundation.

Even if OpenStack emphasizes the joint development of a large and complex open source ecosystem, there are competing firms within its community. For instance, by October 2014, third-party software developers could choose from three different firms offering OpenStack-based cloud computing public cloud services: HP, Canonical and Rackspace. All those three firms were fighting for revenues from OpenStack-based cloud computing services marketed under different brands: *HP Helion Public Cloud*, *Ubuntu Cloud*, *Ubuntu BootStack*, *Rackspace Cloud Servers* and *Rackspace Public Cloud*. In other words, besides contributing to the same project, all the three mentioned firms competed for revenues from third-party actors "renting" public cloud services. Within the hardware business, many contributors to OpenStack such as IBM, HP and Nebula also compete in sales of specialized hardware for cloud-computing installations.

In order to provide evidences of competitive issues, we exemplify in Table 2 and Fig. 1 how OpenStack firms directly compete with each other for the same revenue streams. The network nodes in Fig. 1 represent the top firms contributing to the OpenStack ecosystem while the edges connect firms that compete for the same revenue stream.

## 5 Methodology

We have combined qualitative analysis of archival data (QA), mining software repositories (MSR) and Social Network Analysis (SNA) on publicly-available and naturally-occurring data from the OpenStack Nova repository in order to re-construct and visualize the evolution of collaborations in a sequence of networks. Table 3 presents a set of multidisciplinary methodological notes that guided our mixed methods research design.

We started in a qualitative way, by screening publicly available data such as company announcements, financial reports and specialized press reports, which allowed us to review an immense amount of on-line information pertaining to the competitive cloud computing industry. While taking into consideration established methodological notes that legitimate the use of archival data when studying a case [10, 11, 50–52, 138], we gained valuable insights from the industrial context surrounding the OpenStack community. After attaining a better understanding of the industrial collaborative and competitive dynamics, we extracted and analyzed the social network of the OpenStack Nova project by leveraging SNA [53–55]. By mining digital traces of code collaboration, and by uncovering the social structure of the OpenStack Nova project, the computerized SNA revealed key preliminary understandings of coopetition in the OpenStack ecosystem that were later re-investigated with complementary qualitative data. The combination of methods was not only fundamental for the retrieval of social structures, but also for explaining them.

We first explored social networks data with *Gephi* (v0.8.2) [56] and the *sna* (v2.3-2) and *statnet* (v2014.2.0) statistical modules [57, 58] for *R* (v3.0.2) [59], based on the OpenStack Nova project *changelog*. As in prior multi-disciplinary studies [60, 146–148], our analysis emphasizes the visualization of the collaboration network, which evolves over time, to reveal dynamics among the OpenStack software developers. We then attempted to understand the visualized networks with our acquired understanding from the competitive cloud computing industry in general and OpenStack in particular. The visualization, together with a deeper understanding of the phenomenon under investigation, corresponds to the notion of figuration [61] as pointed out in several studies [62, 63, 147]. Studies which take a SNA perspective for building social network visualizations are already becoming established in Software Engineering studies in general, and in FLOSS research in particular [64–67, 141]. Even so, very few Software Engineering studies exploit the potential of social network visualizations for exploratory research as suggested by [68].

**Table 2** Evidences of coopetition among OpenStack's contributors by revenue model

| Revenue model | Exemplar competitive dyad | Brief description |
|---|---|---|
| Complementary services | HP vs. Mirantis | Both firms compete for Open-Stack related IT projects. Both provide consultancy, integration, customization, testing, deployment, among other IT services. The "body shopping" businesses model is often employed (i.e., the practice of providing technology workers for a contracted short-term project). Price is often determined on a project basis involving long vendor/client negotiations. |
| Complementary software | VMware vs. Cisco | Neutron, the OpenStack cloud networking controller includes a list of plugins that enable interoperability with various commercial and open source network technologies, including routers, switches, virtual switches and software-defined networking (SDN) controllers. Both VMware and Cisco provided such complementary commercial software plugins. By offering this complementary commercial software, VMware also leveraged its visualization technologies while Cisco also leveraged its physical network solutions. |
| Complementary hardware | IBM vs. Intel | OpenStack real-life deployments often require costly hardware capabilities. OpenStack is most often deployed in multiple-processor computer systems at specialized data center facilities. By September 2014, IBM was marketing its new POWER8 CPU architecture as OpenStack-friendly. Meanwhile, Intel marketed its Atom C-2750 and Xeon E5-265x-class processors as optimized of OpenStack deployments. |
| Distribution & support | Red Hat vs. Canonical | Both Red Hat and Canonical designed business models around the commercial support of Linux distributions: e.g., Red Hat Enterprise Linux vs. Ubuntu Advantage. The same applies to OpenStack distributions: By September 2014, Red Hat commercially distributes and supports the Red Hat Enterprise Linux OpenStack Platform vs. Canonical's distribution of Ubuntu OpenStack. Even if OpenStack is freely distributed in an open source way, many enterprise customers opt by a commercially supported distribution with legal contracted service level agreements. |
| Public clouds hosting | HP vs. Rackspace | Similarly to Amazon, HP and Rackspace also provide public cloud services. Unlike Amazon, the offering of HP and Rackspace relies on OpenStack technologies. Any third part actor can contract OpenStack based cloud services both from HP (HP Helion Public Cloud) or Rackspace (Rackspace Public Cloud). The price of cloud computing services are often determined as a function of renting time multiplied by needed capacity (number of CPU nodes, storage and network requirements). |

### 5.1 Data collection
Our screening of public and naturally-occurring data available on the Internet followed established methodological guidelines on case study research [10, 138] which are consistent with more focal and discipline-oriented guidelines established in Software Engineering [11, 69], Information Systems [50, 51], Operations Research [52] and Management [70, 71].

More especially, we have reviewed the most relevant public announcements of companies, publicly available financial reports, publicly available documentation supporting software development, news from both specialized and generalist press, discussions in forums, white-papers and blogs. The selection of sources took in consideration key guidelines on how to conduct qualitative empirical research online [72, 73]. The Kozinets' four criteria for selecting from on-line data sources [72] set the departure points from where we started collecting relevant empirical material.

We also took into consideration specific notes on how to account archival data within a case study, we counteract possible biases by including many and diverse media sources [10, p. 12,13]; we exploited peer debriefing by conducting the analysis as a group instead of one working alone [69, p. 12,13]; and organized the collected quantitative textual data within an digital content management system for meticulous record-keeping [50, p. 374].

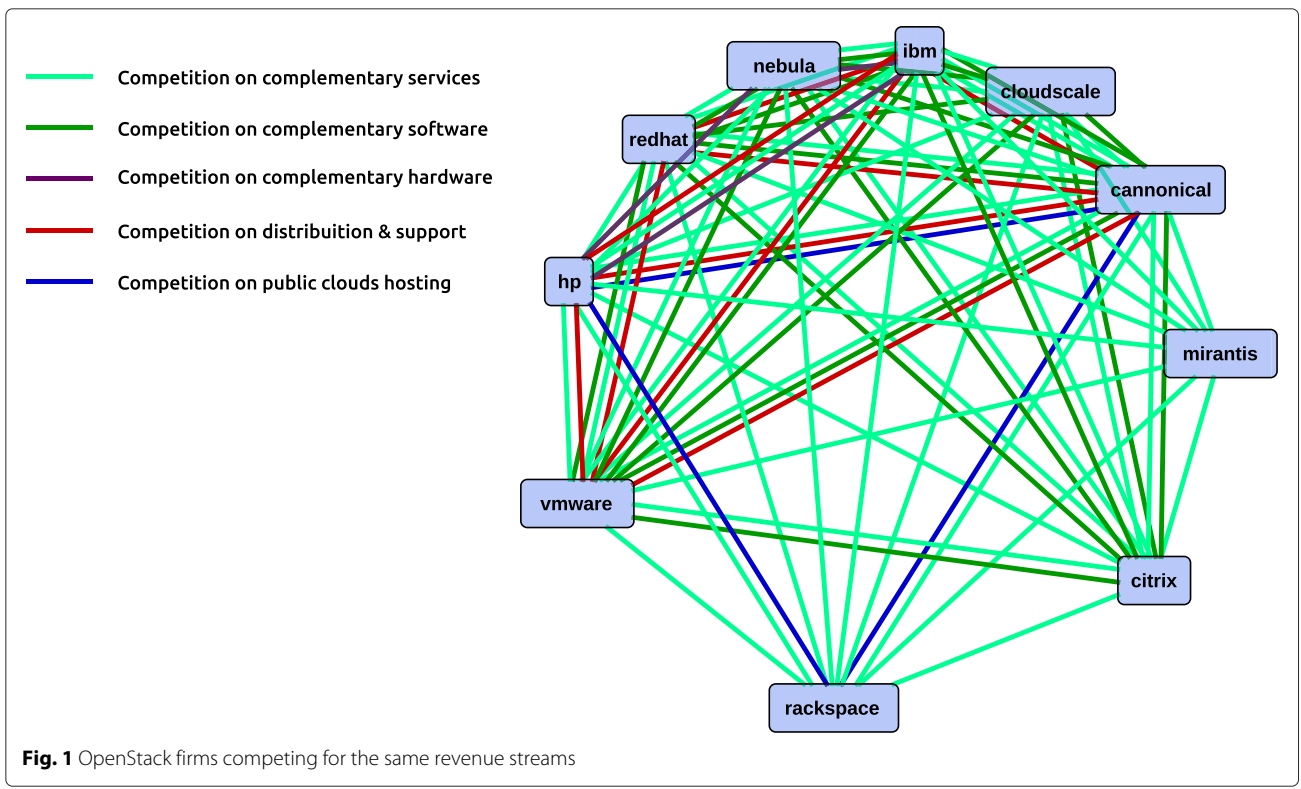

**Fig. 1** OpenStack firms competing for the same revenue streams

**Table 3** Multidisciplinary approach

| Employed approach | Discipline(s) | Seminal works |
|---|---|---|
| Case study rooted on archival data | Multidisciplinary | Yin [10] |
| | | Eisenhardt [138] |
| | | Mockus et al. [139] |
| | | Runeson and Höst [11] |
| Mining software repositories | Software-Engineering | Madey et al. [140] |
| | | López et al. [141] |
| | | Kagdi et al. [142] |
| Network analysis of digital trace data | Software-Engineering | Robles et al. [143] |
| | Information-Systems | Hahn et al. [144] |
| | | Howison et al. [145] |
| Network analysis with emphasis on the visualization of collaborative activities | Biomedicine | Lundvall [146] |
| | Bibliometrics | Cambrosio et al. [147] |
| | Innovation-Studie | Glänzel and Schubert [148] |
| Network analysis of massive networked data. Use of clustering and sub-community detection algorithms. | Physics | Zachary [149] |
| | Mathematics | Kleinberg [150] |
| | Computer-Science | Newman and Girvan [151] |
| | Medicine | Adamcsek et al. [152] |
| | Anthropology | Brohee and van Helden [153] |
| | Neurology | Fortunato [154] |
| | Bioinformatics | Nick et al. [155] |

In Table 4 we point out our selected departure points for collecting relevant on-line archival data. From the initially selected sources, we followed many 'links' and end up visiting many other sites (e.g., corporate sites from companies involved in the development and commercialization of OpenStack technologies or blogs from individual OpenStack contributors).

After the retrieval and qualitative analysis of the selected archival data, we also conducted SNA which allows us to depict overall pictures of the coopetitive dynamics among different developers in the project. Knowledge gained from earlier phases of this study (more qualitative) on OpenStack's actors, events, processes and technology informed the subsequent SNA design.

The input data of SNA is based on different source code release versions of the OpenStack project. The latest source code snapshot from the repository of the OpenStack Nova project was performed on 16 June of 2014, using *git* (v1.8.5.1). From the repository changelog documentation, we extracted basic information, including developer email addresses and the time stamp when changes to a specific file had been made (see Fig. 2). We then connect the developers who work on the same file, and construct a network of collaboration activities among the developers. With the visualization of the collaboration network over time, we aim to understand the evolution of the code-based collaborations with a lens of social structure.

The process of associating the developer email address with code commits is one of the most challenging steps in this research. First, we developed a set of *Python* scripts that validated and corrected repository commit data, for instance to deal with small mistakes performed by developers that submitted code with *Unrecognized author (no email address)*, *Unrecognized author (with email address)* or with an invalid email. Second, we had to deal with the fact that some developers do not commit their changes with their corporate account (e.g., a developer commits code using `dev1@gmail.com` instead of `dev1@hp.com`, while actually affiliated with HP). This problem was tackled by manually triangulating our automatically-retrieved affiliation results with data from the Foundation affiliation database using two external data sources [74, 75].

## 5.2 Data analysis

Our empirical materials span the time period from October 21st 2010 to April 17th 2014. After the `Cactus` release (April 15th 2011), OpenStack abandoned the 3-month time-based release cycle for a coordinated 6-month release cycle with frequent development milestones [76–78]. Thus, we opted to take a longitudinal approach and to construct SNA visualizations that depict collaborative behaviors release after release. Table 5 lists the 9 releases of OpenStack addressed by this study.

The design choice was inferred by characteristics of the OpenStack project: 1) The cyclical nature of OpenStack development where each release cycle encompasses planning, discussion, implementation and fixing release-critical bugs in a sequential way [78]; and 2) The OpenStack scheduling policy, where developers are discouraged from implementing new features during the last development milestone and concentrate efforts on bug-fixing for the coming release candidates [78–80]. As illustrated in Fig. 3, we narrow code-collaboration as a synchronous behavior happening within a release. It is important to remark that due to this design choice our research does not capture collaboration between two developers who contributed to the same file but in a different release cycle.

**Table 4** Internet sources of naturally occurring material

| WWW Internet site | WWW Internet site correspondent title |
| --- | --- |
| http://www.openstack.org/ | OpenStack Open Source Cloud Computing Software |
| http://stackalytics.com/ | Stackalytics | OpenStack community contribution ... |
| http://bitergia.com/ | Software development analytics for Open Source projects .. |
| http://cloudarchitectmusings.com/ | Musings On Cloud Computing and IT-as-a-Service |
| https://www.datacenterdynamics.com/ | Datacenter Dynamics |
| http://slashdot.org/ | Slashdot: News for nerds, stuff that matters |
| http://www.zdnet.com/ | Technology News, Analysis, Comments ... for IT Professionals |
| http://news.cnet.com/ | Technology News - CNET News |
| http://www.computerworld.com/ | IT news, features, blogs, tech reviews, career advice |
| http://techcrunch.com/ | The latest technology news and information on startups |
| http://www.bbc.co.uk/ | British Broadcasting Corporation |
| http://www.nytimes.com/ | The New York Times - Breaking News, World News |
| http://www.todayon-line.com/ | Comprehensive Singapore and international news and analysis |
| http://www.koreatimes.co.kr/ | The Korea Times |
| http://www.nation.co.ke/ | Breaking News, Kenya, Africa, Politics, Business |
| http://elpais.com/ | EL PAÍS: el periódico global |
| http://www.folha.uol.com.br/ | Folha de S.Paulo - Jornal on-line com notícias, fotos e vídeos |

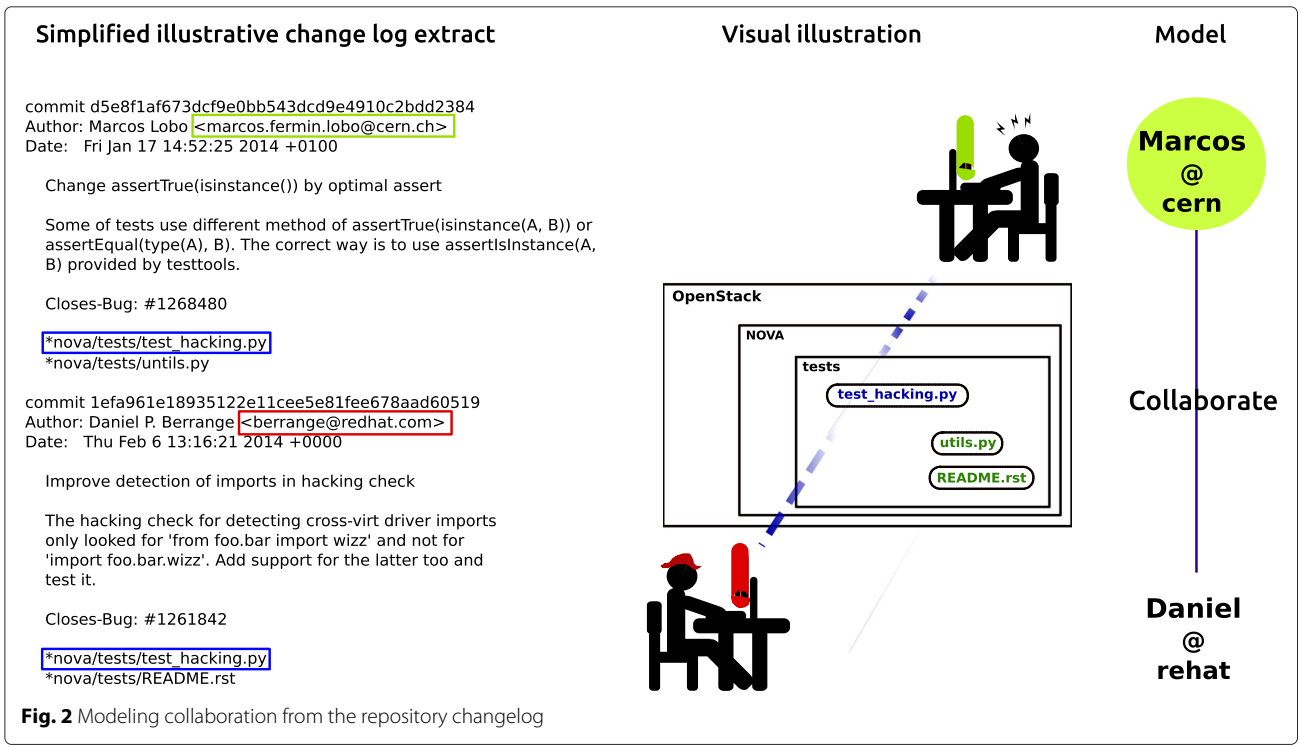

**Fig. 2** Modeling collaboration from the repository changelog

In line with existing guidelines on how to combine digital trace data with SNA [145], we constructed the collaboration network of developers for each partitioned time-slice (i.e., by software release date). In this way, we are able to assess how the collaboration network has evolved over time in response to the exogenous events in the industry.

**Table 5** Releases of the OpenStack community

| Release date | Release name |
| --- | --- |
| Oct 21st, 2010 | Austin |
| Feb 3rd, 2011 | Bexar |
| Apr 15th, 2011 | Cactus |
| Sep 22nd, 2011 | Diablo |
| Apr 5th, 2012 | Essex |
| Sep 27th, 2012 | Folsom |
| Apr 4th, 2013 | Grizzly |
| Oct 17th, 2013 | Havana |
| Apr 17th, 2014 | Icehouse |

During the planning phase of each release, the community gathers for a Design Summit to facilitate live developer working sessions and to assemble the OpenStack roadmap. Documentation regarding the development status of the current release and decisions made at each Design Summit are publicly available [81].

From publicly available data [74, 75] and reports supporting decision-making in the OpenStack ecosystem [82], we observed an *onion* model, as addressed in the open source research literature [64, 83, 84], with a relatively dense core [82]. This indicates that a small set of developers (mostly affiliated to a company) account for most of the development activity. Therefore, in order to better understand collaboration in such a complex collaborative network and to minimize the impact of outliers, we decided to focus our research on the developers affiliated with the top 10 contributing firms to the OpenStack Nova project during the complete period under study. These firms were selected using classical code metrics provided by Bitergia and Mirantis (i.e., number of commits, lines of code and number of completed blueprints).

The selected top 10 firms contributing to OpenStack Nova, briefly introduced in Table 6, can be formally defined as:

$$\begin{aligned}TOPTEN =\{&Canonical, Citrix, Cloudscaling, HP, IBM,\\ &Mirantis, Nebula, Rackspace, VMware, Red Hat\}\end{aligned}$$

(1)

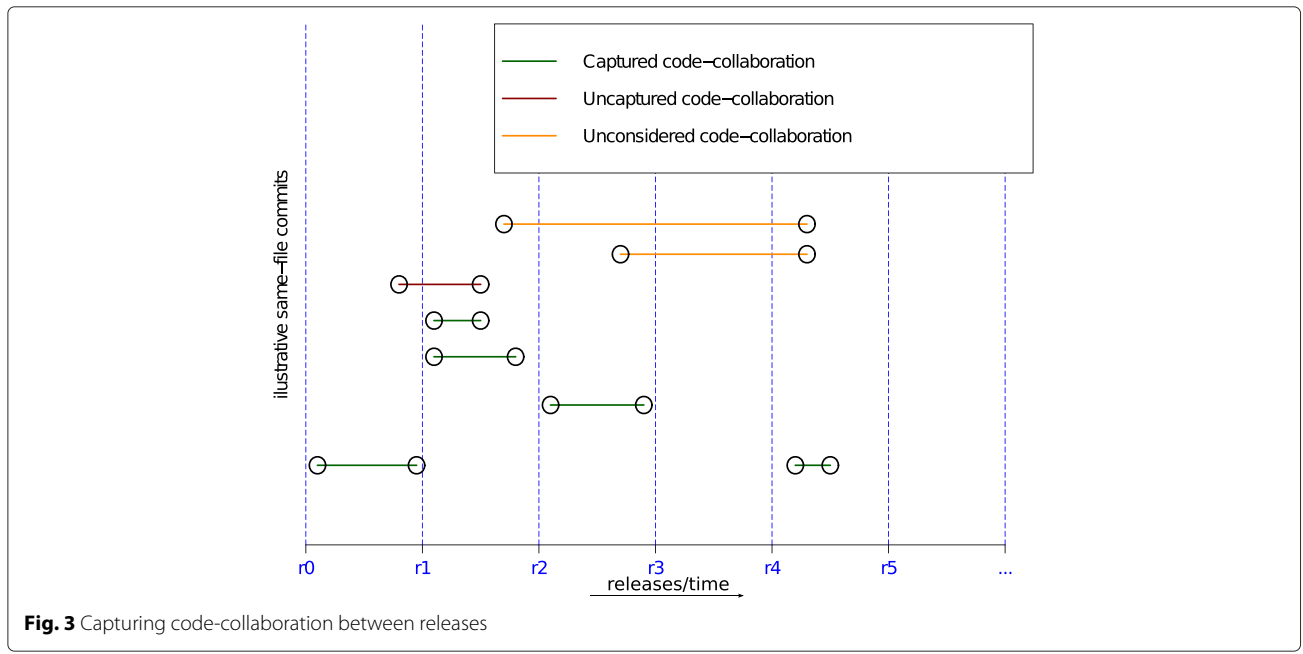

**Fig. 3** Capturing code-collaboration between releases

The collaborative network during a certain time slice can be formally defined as:

$$Gt = (V, Av, E)$$

where:

- $V$ is the set of nodes representing the developers $\in TOPTEN$ contributing to the OpenStack Nova open source software project. All other developers not affiliated with the top 10 contributing firms are not considered in this study.
- $Av$ is the set of nodes-attributes, capturing the company affiliation of a developer.
- $E$ is the set of edges, identifying the connections between two developers if they have worked on the same software source code file. An edge will exist *iff* two developers have modified the same file during the release under study. Edges are both unweighted and undirected.

Various numeric network measures have been established in SNA: for example, eigenvector-centrality [85, 86], degree-centrality and betweenness-centrality [53]; all revealing the importance of a node in a network. Other aspects of a network can also be manifested with important measures such as network-density [54], cluster coefficients [87], strength of ties [88], etc.

However, as our SNA goes hand-in-hand with qualitative analysis of a very competitive and dynamic environment, we concentrated our efforts in visualizing the network graphs, which was sufficient to uncover some history line and to reveal the dynamics of coopetition in an observational way. Although the visualization of social networks has been widely utilized by scholars [60, 64, 141, 147, 148], few studies have explored the time dimension in order to observe how networks evolve longitudinally [3, 89]. At a later stage, we also used another specialized software tool, *Visone*(v2.7.3) [90], to visualize a sequence of networks according to the established release cycle, and to interpret the network evolution with understanding generated from the collection of rich qualitative material capturing the network dynamics.

**Table 6** Top 10 firms contributing to OpenStack

| Firm | Firm description |
|---|---|
| Canonical | The makers of Ubuntu. Provider of support services for Ubuntu deployments in the enterprise. |
| Citrix | Multinational software company that provides virtualization, networking, software-as-a-service (SaaS), and cloud computing technologies. |
| Cloudscaling | Services and open source products company selling custom cloud infrastructure for large service providers, chiefly telecom service providers. |
| HP | Multinational IT company. Provides hardware, software and services to consumers, small- and medium-sized businesses (SMBs) and large enterprises. |
| IBM | Multinational technology and consulting corporation. |
| Mirantis | North California software company specialized on OpenStack. |
| Nebula | North California hardware and software company specialized on cloud computing. |
| Rackspace | Multinational IT hosting company. |
| VMware | Software company that provides cloud and virtualization software and services. |
| Red Hat | Multinational software company providing open source software products to enterprises. |

Given the high density of the collaborative networks during the last releases, and in order to better address **RQ3** (*who tends to collaborate with whom in the project*), we also explored sub-community detection methods. Among a number of possible methods from Graph Theory [91, 92, 153, 154], we opted for a novel technique based on the extraction of Simmelian backbones [155], due to its efficiency to analyze complex networks with unweighted edges.

The obtained social network visualizations (key part of our data analysis) added rigor and comparative logic to the qualitative description (via triangulation of research methods) as suggested by Eisenhardt [93, 138]. However, it also added 'pictures' of the social structure, which *per se* increased the richness of the qualitative description as rejoindered by Dyer [94].

## 6 Results
### 6.1 A overview of the software development process

Directly addressing **RQ1**, we constructed a brief qualitative description of collaborative software development of OpenStack, as can be found in the software engineering literature for other projects [83, 139]. Such description directly derives from many sources of archival data which have been preserved throughout the history of OpenStack [76–80]. The process data is naturally occurring (i.e., not provoked by the researchers), not created for research purposes but to guide and steers real software developers contributing to OpenStack. As our analysis covered more than 4 years of the OpenStack project lifespan, and as the process kept evolving, the description is based on the most recent releases of OpenStack (i.e. years 2014–2015).

OpenStack operates a time-based cyclical software development process where each release cycle encompasses planning, discussion, implementation and fixing release-critical bugs (all in a sequential way). During our investigation we notice that during the earlier release phase, the 'coding' efforts are discussion and specification oriented, while in a later release phase (i.e., stabilization of release candidates) the development turns into bug-fixing mode (as reported in other open source projects [95, 96]).

The 'planning stage' is at the start of a cycle, just after the previous release. This phase usually lasts 4 weeks and runs in parallel with the OpenStack Design Summit on the third week (in a mixture of virtual and face-to-face collaboration). The community discusses among peers gathering feedback and comments. In most cases, specification documents are proposed via a support system [97] that should precisely describe what should be done. Contributors may propose new specs at any moment in the cycle, not just during the planning stage. However doing so during the planning stage is preferred, so that contributors can benefit from the Design Summit discussion and the elected Project Team Leads (PTLs) can include those features into their cycle roadmap. Once a specification is approved by the corresponding project leadership, implementation is tracked in a blueprint [98], where a priority is set and a target milestone is defined, communicating when in the cycle the feature is likely to go live.

The 'implementation stage' is when contributors actually write the code (or produce documentation, test cases among other software related artifacts) mapping the defined blueprints. This phase encompasses a number of milestone iterations (a characteristic of agile software development methods). Once developers perceive their work as ready to be proposed for merging into the master branch, it is pushed to OpenStack's Gerrit review system for public review [99]. It is important to remark that in order to be reviewed in time for a milestone, the change should be proposed in the weeks before the targeted milestone publication date. An open source software collaboration platform [100] is used to track blueprints in the 'implementation stage'. In an open source way, it is worth remarking that not all "features" have to go through the blueprints tracking: contributors are free to submit any *ad hoc* patch. Both specifications and blueprints are tools supporting the discussion, design and progress-tracking of the major features in a release. Even if the *big* corporate contributors are naturally more influential in the election of PTLs steering the tracking process, it should not prevent other contributors from pushing code and fixes into OpenStack. Development milestones are tagged directly on the master branch during a two-day window (typically between the Tuesday and the Thursday of a milestone week).

At the last development milestone OpenStack applies three feature freezes (i.e., *FeatureFreeze*, *DepFreeze* and *StringFreeze*). At this point, the project stops accepting new features and disruptive changes, and concentrates on stabilization, packaging and translation. The project turns then into a 'pre-release stage' (termed as 'release candidates dance' [78]). Contributors are encouraged to turn most of their attention to testing the result of the development effort and fix release-critical bugs. Critical missing features, dubious features and bugs are documented, filed and prioritized. Contributors are advised to turn their heads to the quality of the software and its documentation. The development becomes mainly bug-fixing oriented and a set of norms and tools guide this last product-stabilization phase [101, 102]. Any change proposed for the master branch should at least reference one bug on the bug-tracking system. Once all the release-critical bugs are fixed, OpenStack produces the first release candidate for that project (named RC1).

The OpenStack release team is empowered during this last phase. It creates a `stable/*` branch from the current state of the *master branch* and introduces any new release-critical fix discovered until the release day. Between the RC1 and the final release, OpenStack looks for regression and integration issues. RC1 may be used *as is* for the final release, unless new release-critical issues are found that warrant a RC respinning. If this happens, a new milestone will be open (RC2), with bugs attached to it. Those RC bug fixes need to be merged in the *master branch* before they are allowed to land in the `stable/*` branch. Once all release-critical bugs are fixed, the new RC is published. This process is repeated as many times as necessary before the final release. As it gets closer to the final release date, to avoid introducing last-minute regressions, the release team limits the number of changes and their impact: only extremely-critical and non-invasive bug fixes can get merged. All the other bugs are documented as known issues in the Release Notes instead.

On the release day, the last published Release Candidate of each integrated project is collected and the result is published collectively as the OpenStack release for this cycle. OpenStack should by then be stable enough for real industrial deployments. But once the version is released, a new cycle will commence within OpenStack; the *master branch* switches to the next development cycle, new features can be freely merged again, and the process starts again.

### 6.2 Longitudinal description of collaborative and competitive issues

To answer **RQ2** we will take a chronological approach. We will proceed by quoting the words of Jim Curry in one of the first announcements of the OpenStack project. The founding leader of the OpenStack community starts by advocating the freedom of open source software before stating the mission of the the OpenStack project. It is important to notice that Jim Curry emphasized the roles of NASA and Rackspace as initial contributors to the project; project did not start from zero.

"What is OpenStack? Well, our mission statement says this:

*To produce the ubiquitous Open Source Cloud Computing platform that will meet the needs of public and private clouds regardless of size, by being simple to implement and massively scalable.*

That is a big ambition. The good news is that OpenStack is starting with code contributions from two organizations that know how to build and run massively scalable clouds - Rackspace and NASA. Rackspace has been in the cloud business for 4 years and now serves tens of thousands of customers on its cloud platform. Likewise, NASA began building their Nebula cloud platform 2 years ago to meet the needs of their scientific community" — Jim Curry, OpenStack Lead, 19 July 2010 [103]

Visualizations in Figs. 4, 5, 6, 7, 8, 9, 10 and 11 provide an understanding of how key players in the cloud computing industry collaborate in a Free/Libre/Open Source software ecosystem. SNA visualizations are not just 'pretty pictures' as network studies are periodically criticized (counter-arguments to this criticism can be found in [104]), but exploit combinations of text, signs, color, size, location and shape to communicate several pieces of numerical information simultaneously [105]. Each network visualization aggregates both numbers, mathematical formulae, and written text which have to be otherwise communicated sequentially [106].

In our visualizations, the size of a node is dependent on its degree-centrality; i.e., the larger the node, the more social connections the developer has. The value of degree-centrality depends on the number of adjacent nodes with which a node is connected. Therefore, the higher a developer's degree-centrality, the more likely he/she is collaborating with others.

Figure 4 captures collaboration in the OpenStack Nova project from the `Austin` to the `Bexar` release, from October 21st 2010 to February 3rd 2011. From it, we can derive the collaboration between software developers affiliated with companies; so, Citrix had three developers working on the project together with Rackspace.

"OpenStack provides a solid foundation for promoting the emergence of cloud standards and interoperability." …. "As a longtime technology partner with Rackspace, Citrix will collaborate closely with the community to provide full support for the XenServer platform and our other cloud-enabling products." — Peter Levine, SVP and GM, Citrix, 19 July 2010 [107]

Citrix who have been working before with Rackspace, wanted to make sure that their XenServer platforms

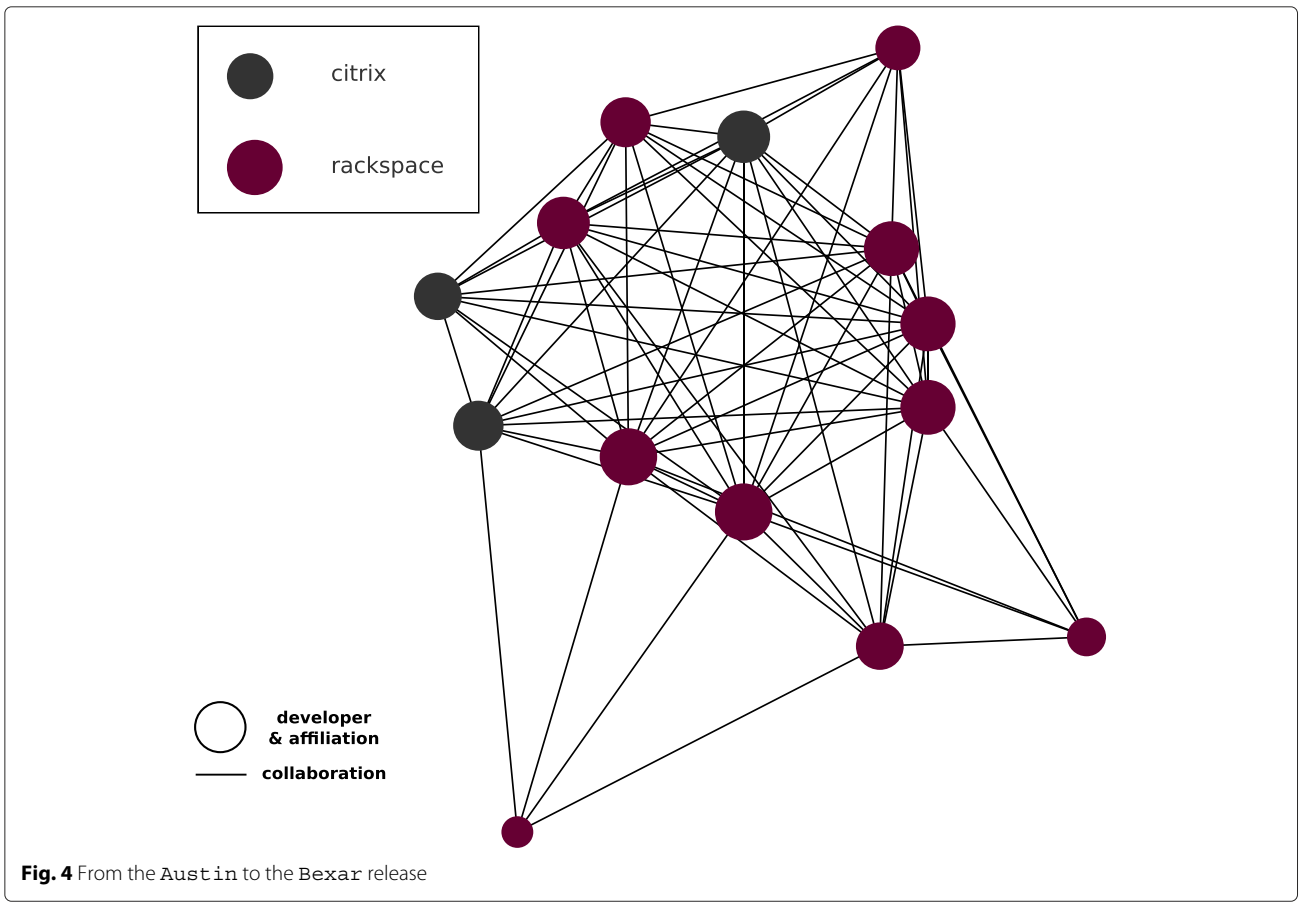

**Fig. 4** From the `Austin` to the `Bexar` release

would be conveniently integrated with Rackspace plans for OpenStack.

> "The project is exhibiting the key benefits that the industry derives from successful open source collaboration: rapid development, faster testing, feedback and project turn around, broader industry adoption and learning through implementation and de-facto standardization whilst avoiding the prospect of commoditization.
>
> It has been rewarding to work with the OpenStack crew, and to have experienced first hand the dedication to an open, code-rules, community-first approach taken by the project leaders. OpenStack has shown that it is possible to rally the community around the development of "management" software - as opposed to the Linux kernel or Xen - and it is definitely the case that OpenStack is breaking new ground for the industry at large. With the release behind us, our team will head in force to San Antonio for the next Design Summit."
> — Simon Crosby, CTO, Citrix 21 October 2010 [108]

Our second visualization with degree-centrality, in Fig. 5, captures collaboration from the `Bexar` to the `Cactus` release (from February 3rd 2011 to April 15th 2011). From this visualization we can observe a new node, a developer affiliated with Cloudscaling. Cloudscaling was founded in 2006 by the cloud architect and open source software advocate Randy Bias, and the co-founder Adam Waters. It started as a professional services company selling custom cloud infrastructure for large service providers, chiefly telecom service providers. They had KT (formerly Korea Telecom) as an early customer, for which the company in 2010 designed and deployed the first OpenStack-based storage cloud outside Rackspace.

> "Earlier this week, one of our clients, a Tier 1 ISP, launched an object storage cloud based on OpenStack, an open source compute and storage framework created by Rackspace and NASA. The new storage cloud is the first commercial OpenStack-based storage offering in the market after Rackspace itself, which is based on the same technology.
> Cloudscaling assisted in developing
> this solution for the new product, including hardware, networking, configuration, systems integration, monitoring and management." – Joe Arnold, Director of engineering, Cloudscaling, 31 of January 2011 [109]

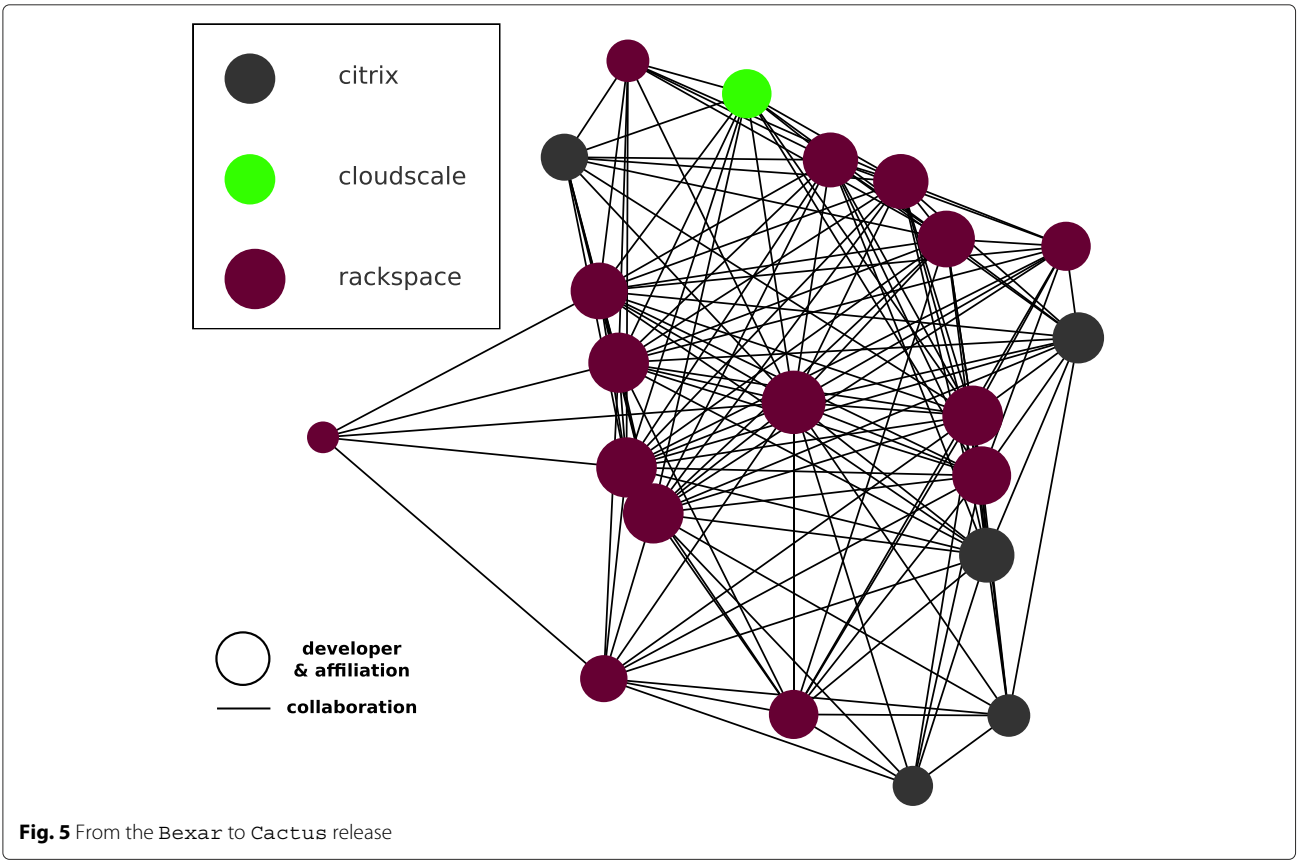

**Fig. 5** From the `Bexar` to `Cactus` release

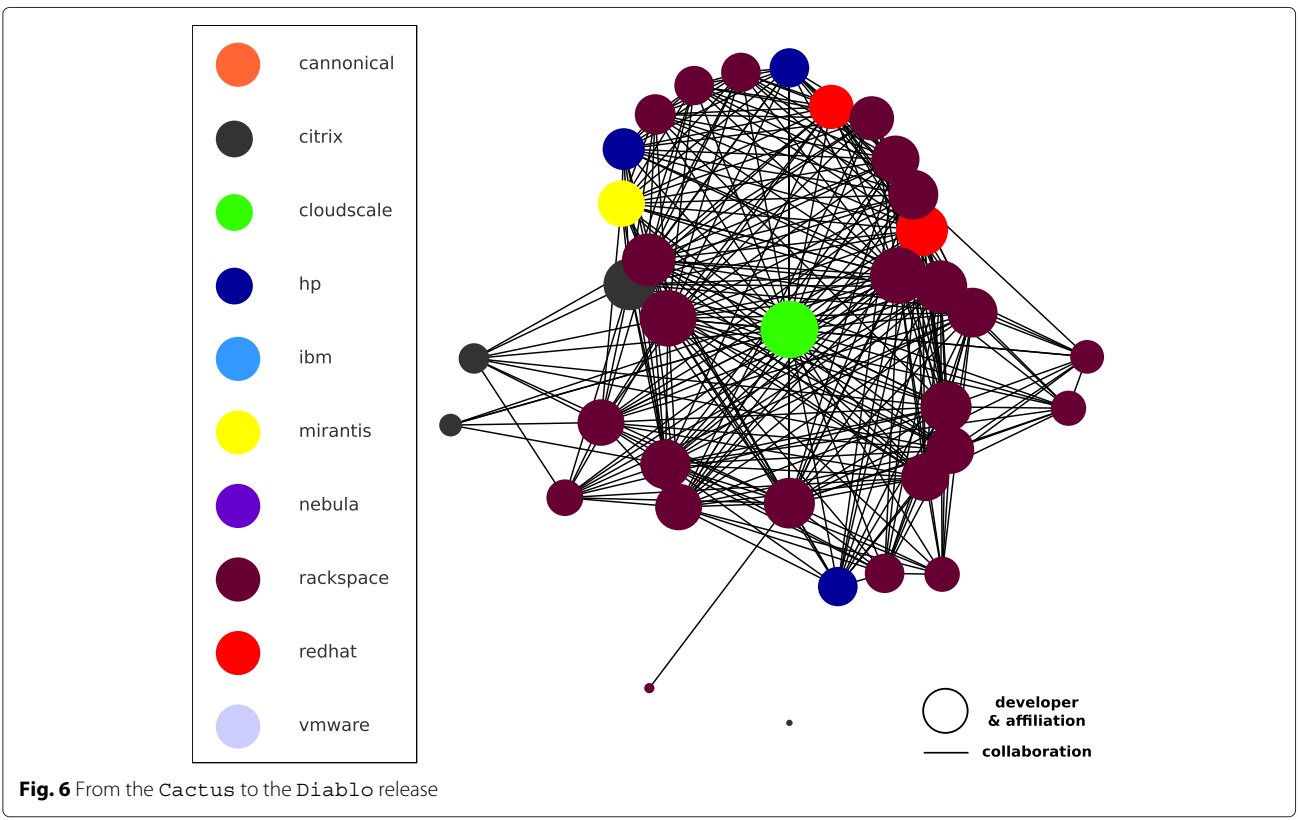

**Fig. 6** From the `Cactus` to the `Diablo` release

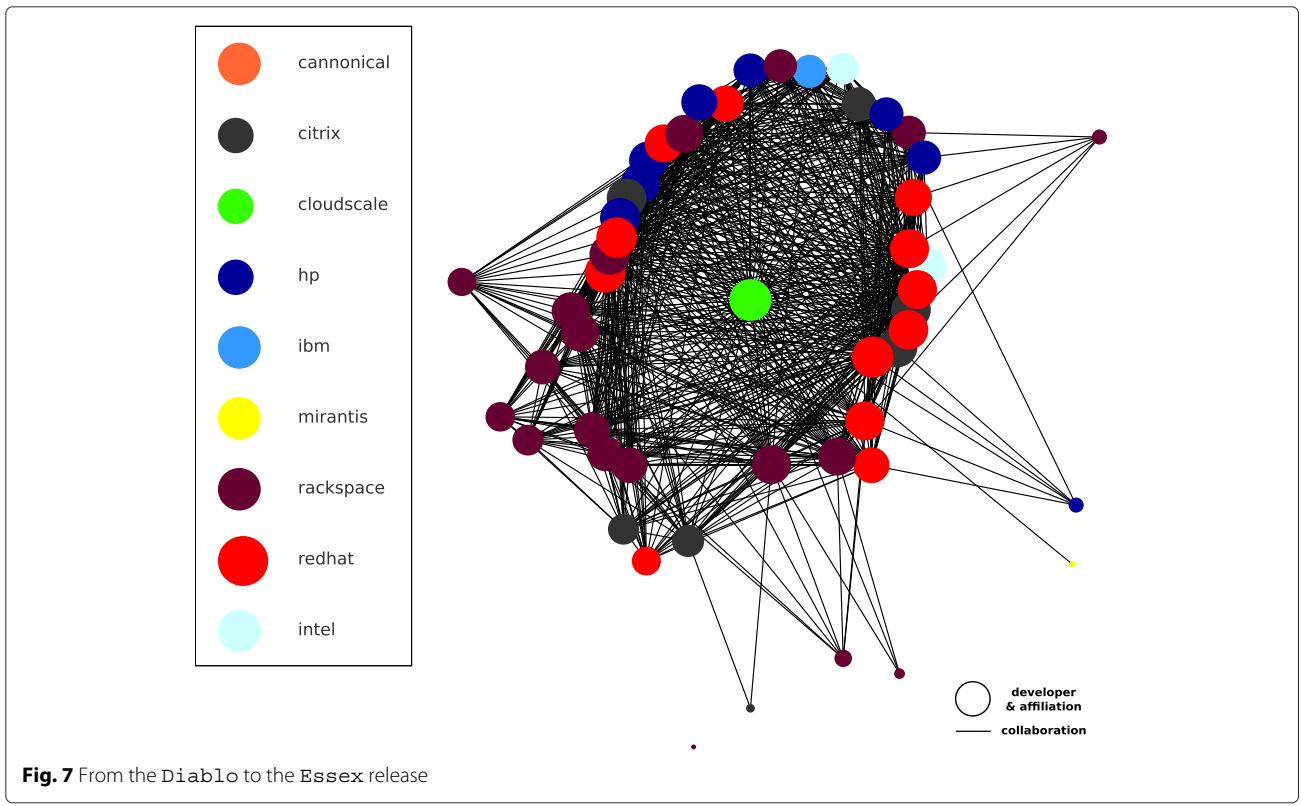

**Fig. 7** From the `Diablo` to the `Essex` release

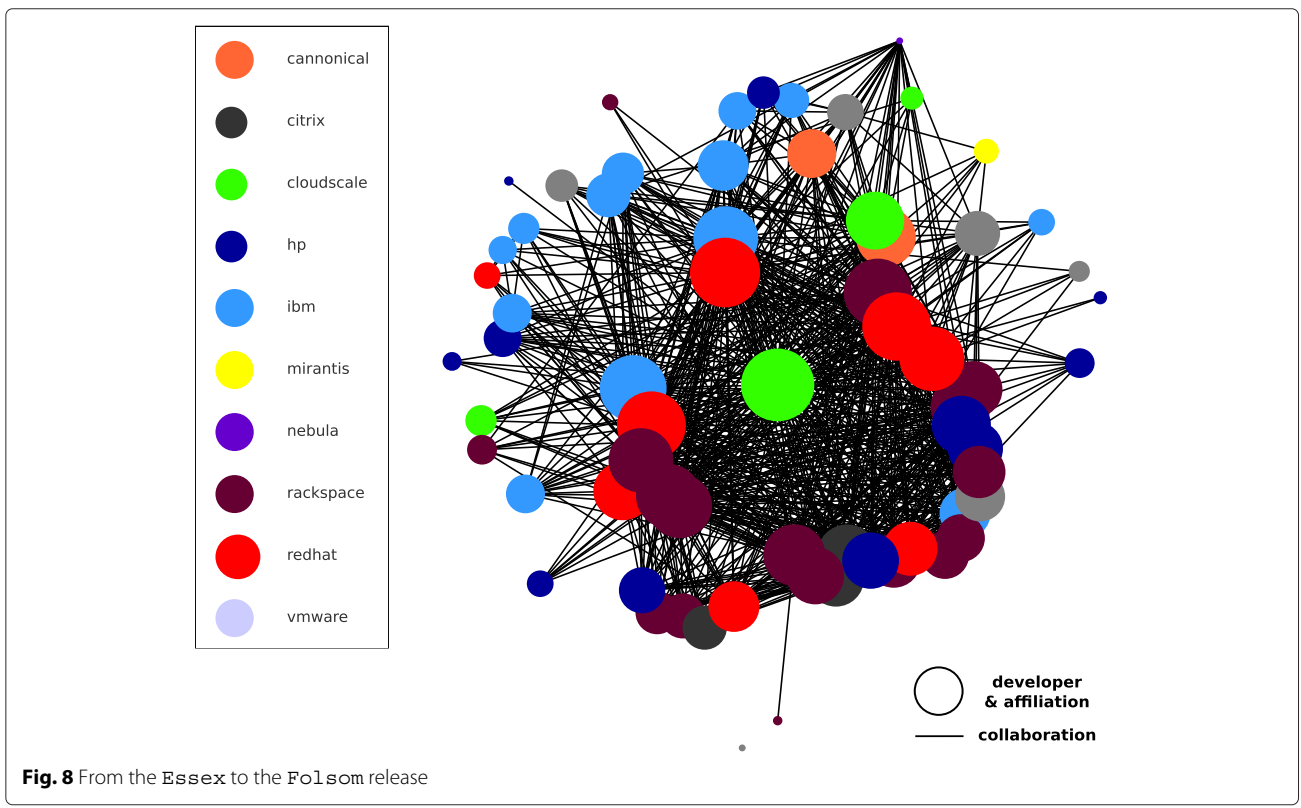

**Fig. 8** From the `Essex` to the `Folsom` release

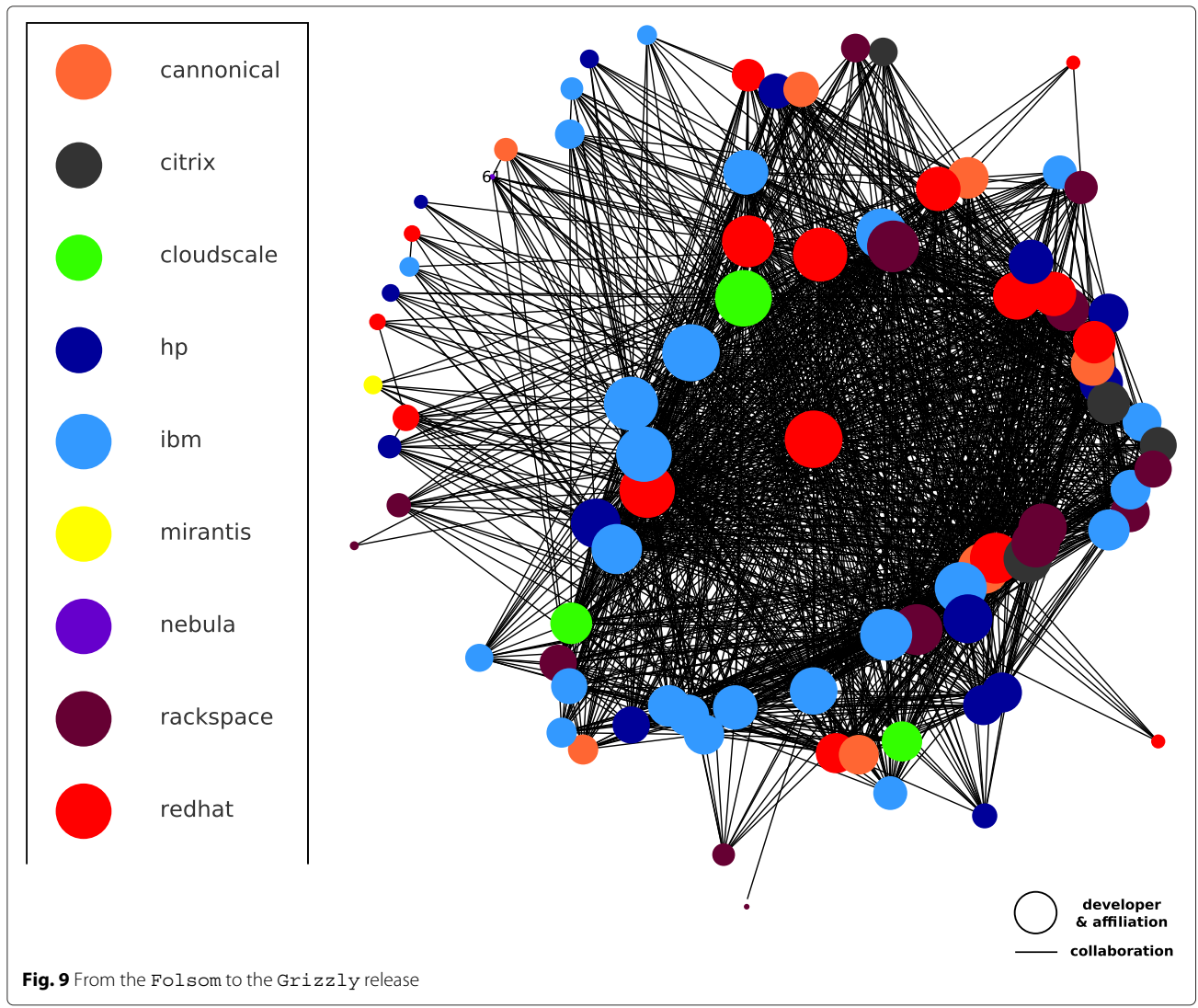

**Fig. 9** From the `Folsom` to the `Grizzly` release

Our visualization in Fig. 6 captures collaboration from the `Cactus` to the `Diablo` release (from April 15th 2011 to September 22nd 2011). HP (a well-known IT multinational company), Mirantis (an OpenStack startup), and Red Hat (the company behind the Red Hat Enterprise Linux and sponsor of the Fedora Linux distributions) joined the coopetitive software development efforts.

Mirantis was founded in January 2011 by Boris Renski Jr. and Alex Freedland. Also born in Northern California, this startup marketed itself as a "pure-play" OpenStack company and started working early with Red Hat. During our qualitative analysis of on-line data on the Internet, we came across a conversation between engineers from Mirantis and an open source enthusiast contributing occasionally to the Fedora project. Such conversation provided qualitative digital trace data evidencing collaboration between Mirantis and Red Hat [110].

"Our internal infrastructure is running on Fedora, instead of migrating the full infrastructure to Ubuntu, we decided to make OpenStack Fedora-friendly." – Maxim Lvov, Senior deployment engineer, Mirantis, 18 of May 2011 [110]

"Are you aware of the upstream effort to create packages for Fedora?" ... "would you be willing to contribute your specs if you really build your rpms from the sources?" – Fabian Deutsch, Contributor to the Fedora project, 20 of May 2011 [110]

"I've had a conversation with David Nalley about contributing to Fedora. Sure, we are willing to contribute. We are under refactoring, and we'll show them soon." – Mike Scherbakov, OpenStack architect, 20 of May 2011 [110]

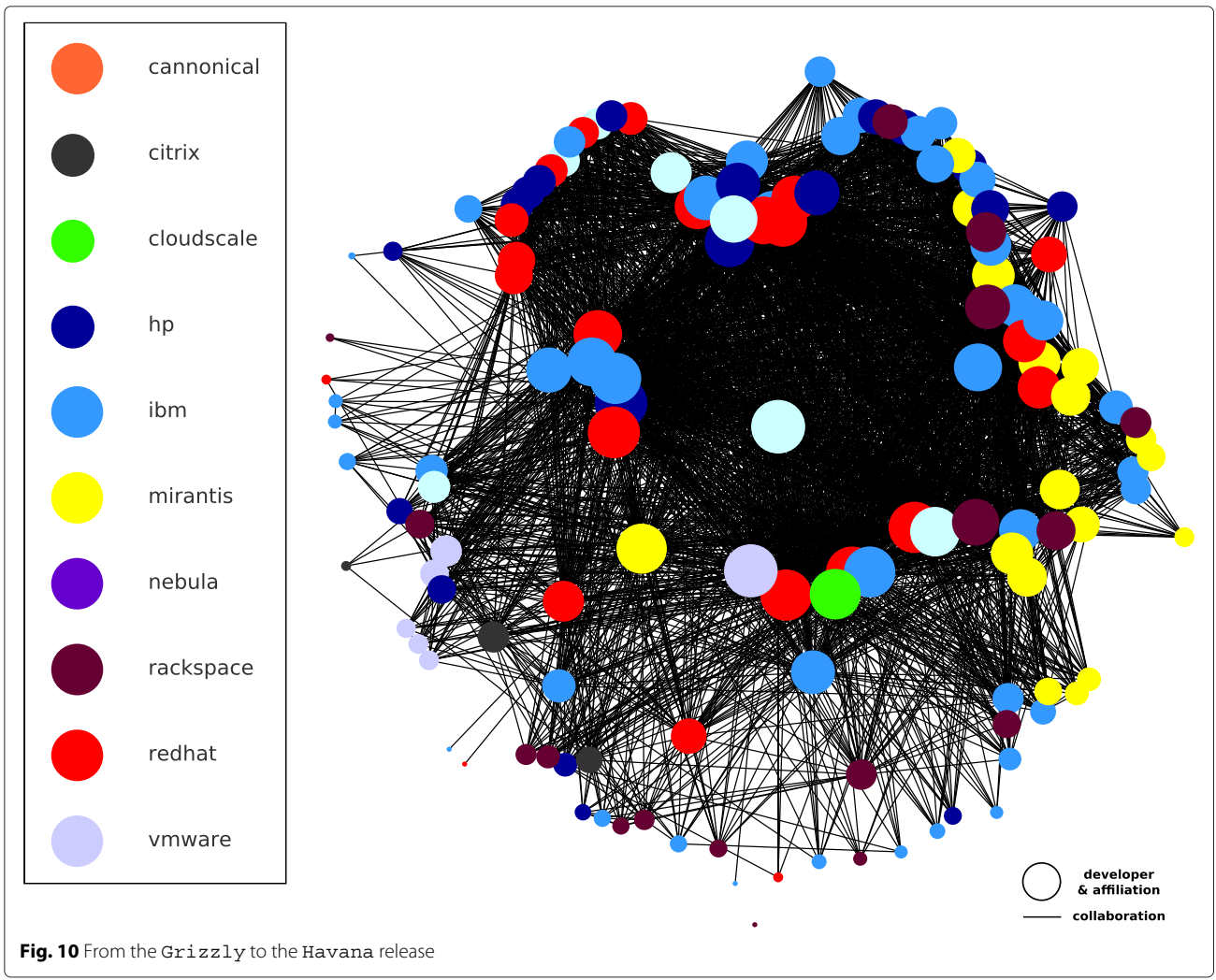

**Fig. 10** From the `Grizzly` to the `Havana` release

Meanwhile, HP started marketing their cloud computing services based on OpenStack. HP markets itself as the leading corporation behind the project, emphasizing that OpenStack is the only cloud computing solution without a single-vendor lock-in, but with an extensive ecosystem behind it [111]. Another key startup to OpenStack was born in those days, Nebula. Nebula was founded also in Northern California in Spring 2011 by Chris C. Kemp, a former NASA Chief Technology Officer, his longtime colleague Devin Carlen, and the entrepreneur Steve O'Hara. The company absorbed much of the original NASA team which coded the first bits of OpenStack [112].

Figure 7 depicts collaborations from the `Diablo` to the `Essex` release (from September 22nd 2011 to April 5th 2012). Although the graph becomes more dense, we can visualize new nodes representing early contributions of Intel (interested in making OpenStack deployments work well on Intel micro-processors) and IBM. IBM has a long history of working with open standards and open source initiatives such as in the Apache and Eclipse projects, and has been able to sell complementary solutions (i.e., hardware, software and services) from open source projects. It expects the same business model to work well with OpenStack.

"Our goal is to accelerate the rate and pace of both functional and non-functional (performance, scalability, reliability, etc.) enhancements to the OpenStack code base. In that vein, IBM will be a very active participant in the next OpenStack Design Summit scheduled for October 15–19 in San Diego. The time has come to establish a de-facto base implementation for IaaS and related open interfaces. Without this, the industry risks fragmentation and complexity that will only serve to slow down the adoption of cloud technology and innovation. Support for OpenStack and the OpenStack Foundation is an effective way to achieve this goal. In a Wired.com blog

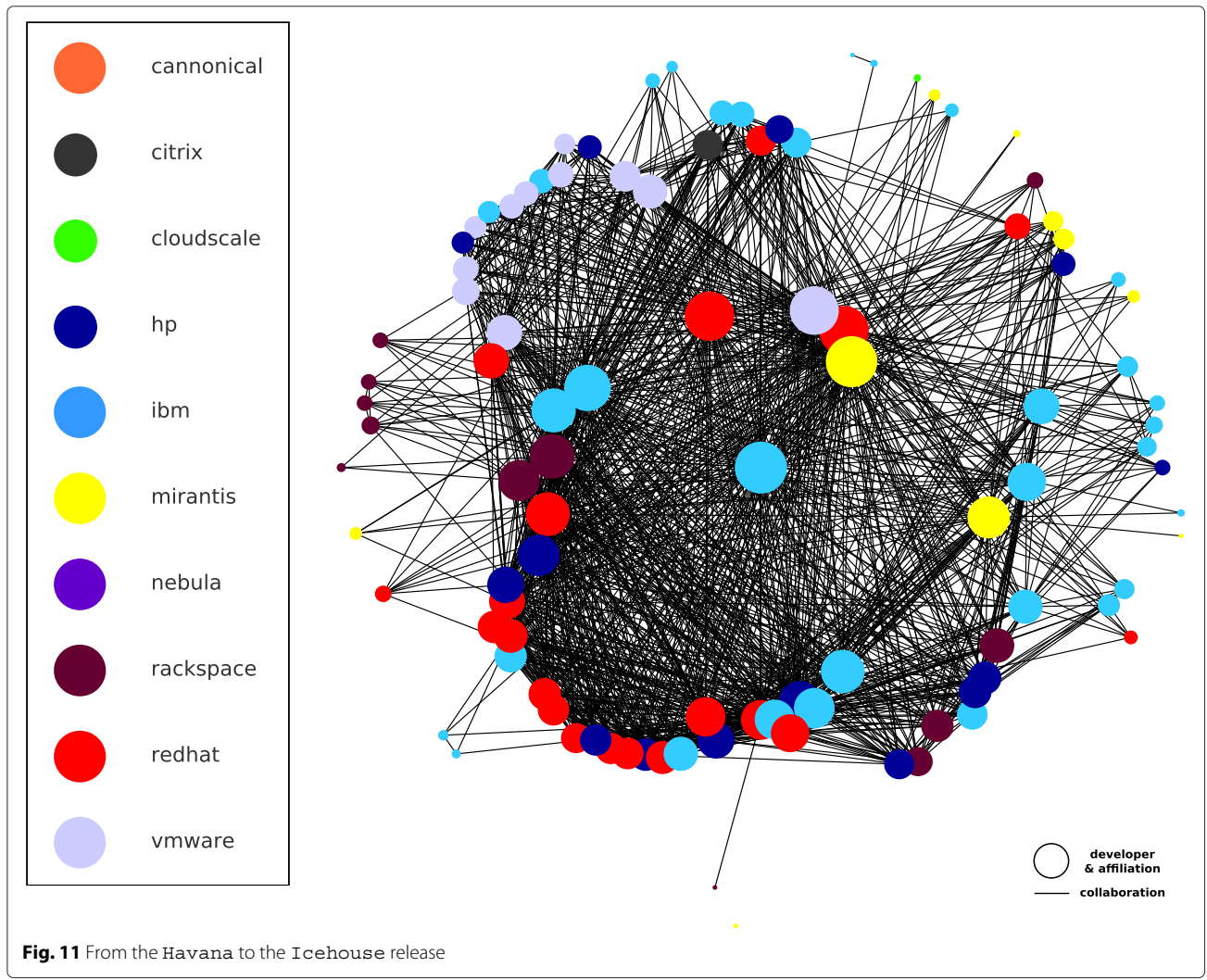

**Fig. 11** From the `Havana` to the `Icehouse` release

I wrote back in April, I highlighted three initial focus areas for IBM: 1) Establish the OpenStack Foundation, 2) Support and expand the OpenStack Ecosystem and 3) Contribute to the OpenStack Development." – Angel Diaz, Vice President of Open Standards, IBM, 19 of September 2012 [113]

Figure 8 shows collaborations between the `Essex` and the `Folsom` release (from April 5th 2012 to September 27th 2012). We can observe that the network becomes more dense, as there are more developers working with each other. Even if some of their developers continued contributing to the project, Citrix had by then abandoned its Olympus OpenStack distribution in order to focus instead on the competing CloudStack cloud computing FLOSS ecosystem. Citrix decided to contribute to the competing CloudStack software ecosystem under the umbrella of the Apache Software Foundation, with a codebase resulting from the acquisition of Cloud.com in July 2011. This turn of strategy from Citrix is related with the OpenStack lack of integration with the Amazon's APIs (Application Programming Interfaces). Amazon is currently the leader of cloud services, and the migration costs to another cloud computing infrastructure are very high, specially if the APIs do not resemble each other.

"Amazon has in many ways invented and created this market, and with what is projected to be $1 billion in ecosystem and customer revenue attached to Amazon cloud, we believe the winning cloud platform will have to have a high degree of interoperability with Amazon" – Sameer Dholakia, GM Cloud Platforms Group, Citrix, 3 of April 2012 [114]

Figure 9 shows collaborations in the OpenStack Nova project from the `Folsom` to the `Grizzly` release (from September 27th 2012 to April 4th 2013). As expected, Citrix reduced its commitment to OpenStack as we observe little activity from Citrix developers. Canonical continued investing increasingly in the development of OpenStack, interested in keeping its Linux Distribution *Ubuntu* as the leading Linux distribution for OpenStack clouds [115].

VMware, a Northern Californian firm with expertise in virtualization technologies, made substantial contributions (evidenced by the number of commits) during this between-releases period [116]. The acquisition of the networking virtualization startup Nicira in July 2012 reshaped the VMware cloud computing strategy. As a sign of commitment to OpenStack, VMware and Canonical issued a joint statement on their intentions to work together to improve the integration VMware technologies with Canonical's OpenStack distribution.

> "(OpenStack Summit) Canonical and VMware, Inc. (NYSE: VMW), the global leader in virtualization and cloud infrastructure, today announced a collaboration that will enable organizations to deploy VMware technologies, including VMware vSphere and Nicira NVP, with Canonical's OpenStack distribution. Canonical's Ubuntu Cloud Infrastructure, the most widely used OpenStack distribution, will now include the plugins required to use OpenStack with vSphere and NVP. Canonical will provide commercial support for OpenStack and will collaborate with VMware on issues related to vSphere or NVP running with OpenStack. In addition, VMware reaffirms its support of Ubuntu as a fully supported guest operating system (OS) on vSphere. This agreement will enable customers the flexibility to deploy and reliably run OpenStack clouds with Ubuntu Cloud Infrastructure on VMware vSphere while receiving commercial support." – Joint press release from VMware and Canonical, Acquire Media, 16 of April 2013 [116]

Figure 10 captures collaboration in the project in a more recent phase, from the `Grizzly` to the `Havana` release (from April 4th 2013 to October 17th 2013). We can see that VMware took its commitment to OpenStack seriously, as six new developers engaged in developing with other OpenStack developers affiliated with *TOPTEN* firms. Mirantis, in yellow on the right of Fig. 10, invested heavily in collaborative activities with IBM, Rackspace and Red Hat. Mirantis counted on financial support from Dell Ventures and Intel Capital (representing the interests of hardware manufacturers betting on OpenStack) [117] and additional investment by Ericsson, Red Hat, and SAP Ventures [118], turning it into one of the biggest code contributors to the OpenStack software ecosystem in just a few months as reported by Bitergia [82]; the number of developers from Mirantis increased from 1 to 17 in this time period.

Figure 11 captures collaborations in the latest period studied, from the `Havana` to the `Icehouse` release (from October 17th 2013 to April 17th 2014). In this visualization, the number of network nodes (i.e., software developers affiliated with *TOPTEN* firms in the OpenStack Nova) decreased, while retaining a similar density. This fact must be interpreted carefully, as it does not mean that the number of software developers contributing to the OpenStack Nova is now lower. There were currently 313 developers contributing with code, and 483 developers reviewing code during this period. The community contributing to the OpenStack Nova project increased significantly, while the role of the *TOPTEN* firms decreased. As of this release, Intel, NEC, Huawei and the rest of the non-affiliated developers grouped together would belong to the TOP 10 contributors of the `Icehouse` release – although considering the whole lifespan of OpenStack the project, we have considered them to be out of the TOP 10.

By this time the role of NASA on OpenStack had diminished. The first developments of OpenStack were in the service of science, supporting NASA's research activities. NASA's prestige and participation has been a selling point for advocates of OpenStack technologies. NASA lost much of its IT staff working on its Nebula cloud computing project. Software developers and IT architects headed to startups and high-tech giants within the OpenStack ecosystem. Moreover, a cost-driven IT reform led to disinvestment in OpenStack by NASA. Today, scientists at NASA depend on Amazon EC2 and Microsoft Azure cloud computing infrastructures [119]. While much of the scientific data Meanwhile, on the other side of the Atlantic, the European Organization for Nuclear Research (CERN) decided on an OpenStack based strategy in 2012. In January 2014, OpenStack was already running collision reconstructions at the Large Hadron Collider (LHC) [120].

### 6.3 Ecosystem and its sub-communities

In order to carefully address **RQ3** we employed sub-community detection methods to discover sub-communities in the ecosystem. Particularly for the last releases, with very dense networks, the direct interpretation of the visualizations is extremely difficult. We opted to use data from the last OpenStack releases (`Grizzly`, `Havana` and `Icehouse`) because of higher project maturity and a steady diminution of group cohesion (i.e., tendency for subgrouping) as "plotted" in Fig. 12. The figure includes three basic social network metrics that capture the evolution of the collaborative network over time/release: number of nodes, number of edges and network density, all correspondent measures of community-size, collaborative behavior and community cohesion.

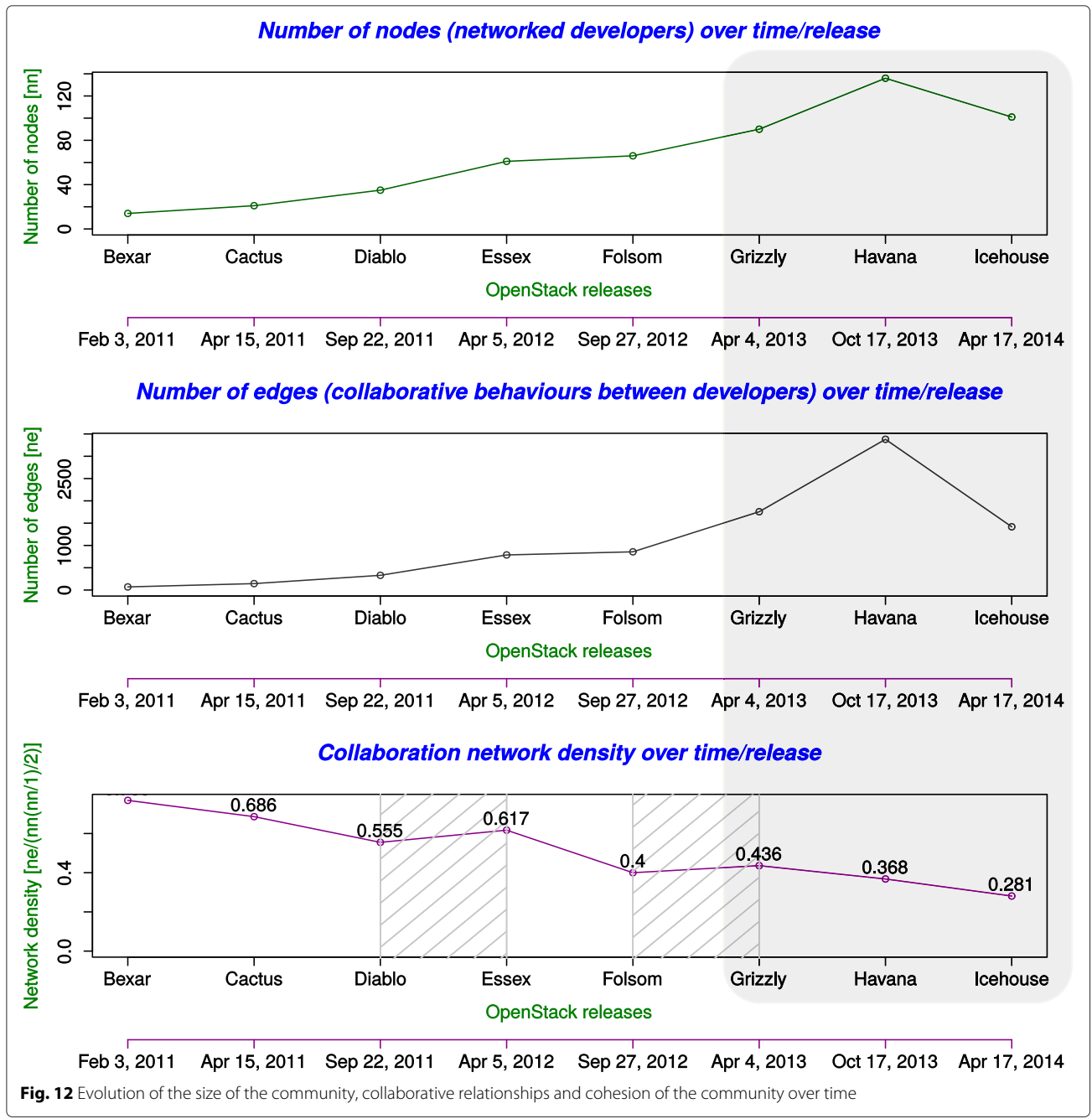

**Fig. 12** Evolution of the size of the community, collaborative relationships and cohesion of the community over time

As a result of the extraction of Simmelian backbones from the collaborative network of the last OpenStack releases, the emergent sub-communities in Fig. 13 reveal, contrary to what the authors expected, a low degree of homophily in code collaboration, meaning that developers do not tend to work with developers from their own company.

Instead of visualizing the complex "raw" collaborative network (see Fig. 13a), by using the sub-community detection technique, we can depict three ecosystem-groups (i.e., sub-communities): a larger one involving IBM, HP, Rackspace, Red Hat and Mirantis (see top-left Fig. 13b); and two smaller ecosystem-groups including 1) VMware, Red Hat, IBM and HP (bottom of Fig. 13b); and 2) IBM, HP and Rackspace and Red Hat (top-right of fig. 13b). There were overlapping communities (e.g., Red Hat is present in all the detected groups). It is also important to notice that among the top 10 firms contributing to the OpenStack project, no firm had *their* developers working with each other in an isolated sub-community.

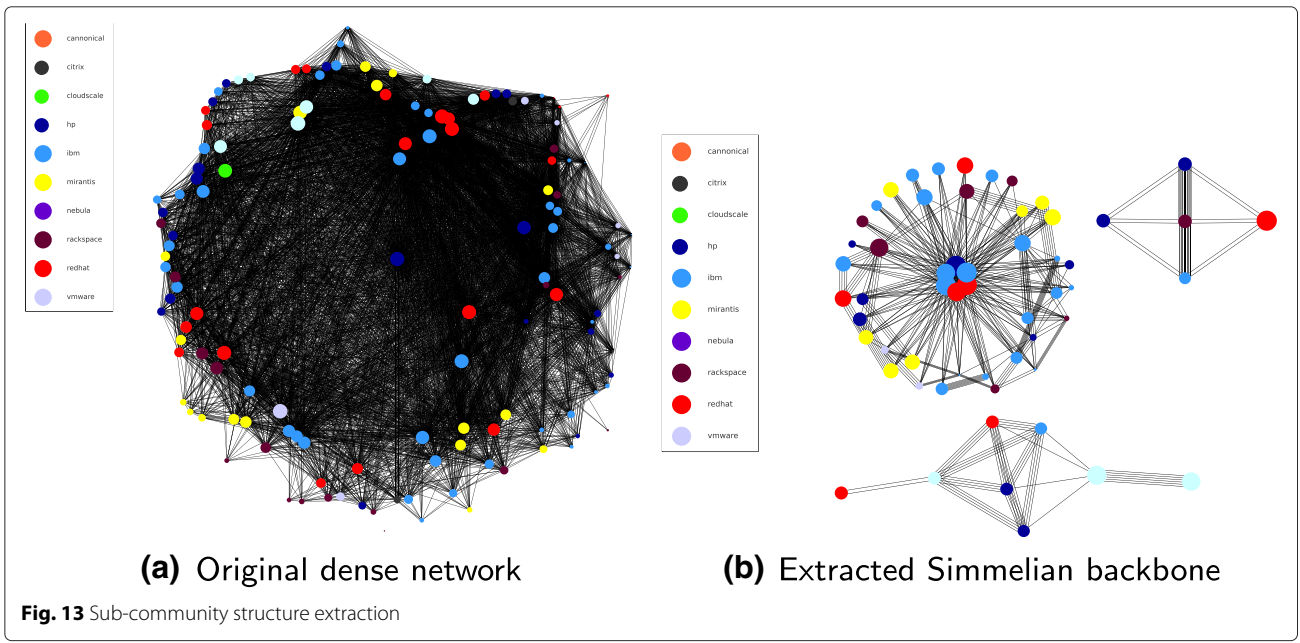

**Fig. 13** Sub-community structure extraction

### 6.4 Exploring the relationship between revenue-models and collaboration

For addressing **RQ4**, we can also exploit network analysis to explore how competition for a revenue stream affects collaboration in the OpenStack Nova case. This can be done (as displayed in Table 7) by calculating the density of the two collaborative networks: collaboration among firms *competing* for a specific revenue stream vs. collaboration among firms that *do not compete* for a specific revenue stream. It should be noted that density is a measure of cohesion, collaboration and learning. Alternatively, but less precisely, it can be also be done by zooming and visualizing side by side collaboration among firms competing vs. collaboration among firms that do not compete (see Fig. 14).

Either by the calculated network densities or by the visualizations, we can conclude the following:

- Companies providing complementary software collaborate **more** than companies not providing complementary software.
- Companies providing complementary hardware collaborate **more** than companies not providing complementary hardware.
- Companies providing distribution and support collaborate **more** than companies not providing complementary distribution and support.
- Companies providing public clouds collaborate **less** than companies not providing public clouds.

**Table 7** Community cohesion across firms competing for the same revenue stream

| Competing revenue stream | Revenue stream description | Competing firms | DO $\frac{n(\alpha_i)}{den(\alpha_i)}$ | DO NOT $\frac{n(\beta_i)}{den(\beta_i)}$ |
|---|---|---|---|---|
| Complementary services | Firms providing specialized support, maintenance, integration, customization, testing, consulting, etc. | ALL TOPTEN | $\frac{136}{1}$ | $\frac{0}{UND}$ |
| Complementary software | Firms providing software embedding OpenStack or complementary drivers and plug-ins. | IBM, VMware, Nebula, Cloudscaling, Red Hat, Citrix, Canonical | $\frac{76}{0.472}$ | $\frac{65}{0.362}$ |
| Complementary hardware | Firms providing hardware targeting OpenStack installations. | IBM, HP, Nebula | $\frac{69}{0.459}$ | $\frac{67}{0.434}$ |
| Distribution and support | Firms distributing OpenStack with enterprise support (for a fee). | Red Hat, IBM, HP, VMware, Canonical | $\frac{91}{0.460}$ | $\frac{45}{0.425}$ |
| Public cloud services | Firms providing OpenStack based public cloud services such as hosting. | HP, Rackspace, Canonical | $\frac{41}{0.3200}$ | $\frac{95}{0.458}$ |

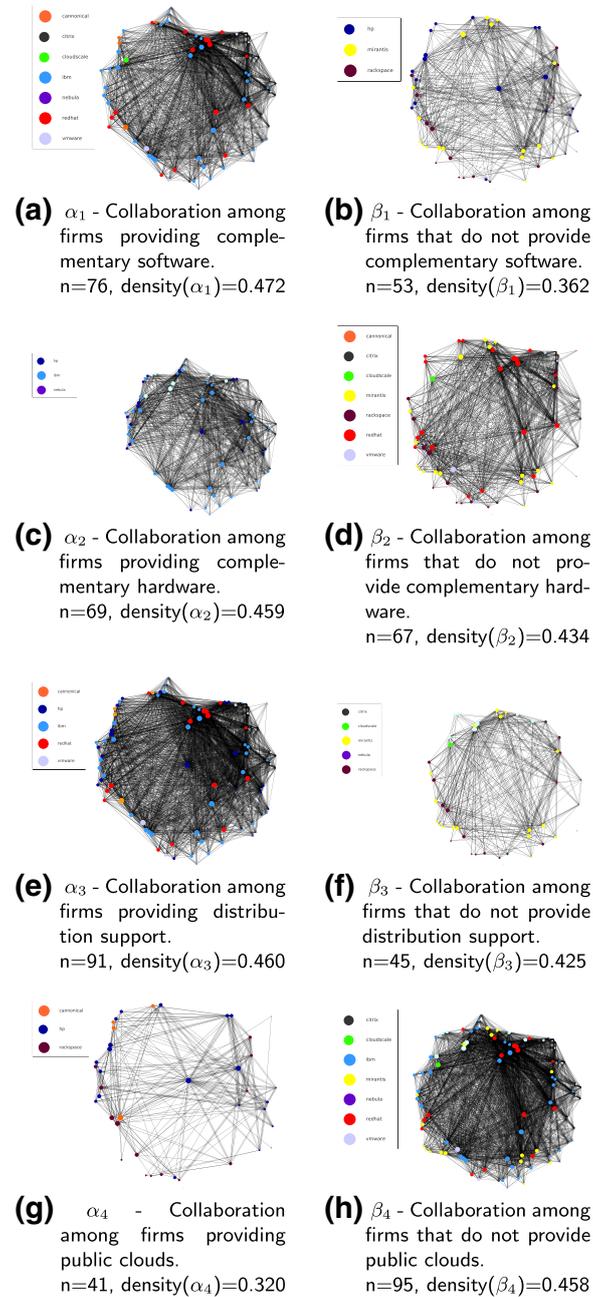

**Fig. 14** Visualizing the impact of competition for the same revenue revenue model

These results suggest that the open competition strategy is working well for the firms involved in the development of the OpenStack Nova. Even if they are competitors for the same revenue stream they do not collaborate less, with the exception of HP, Rackspace and Canonical, who are providers of public cloud computing services.

During our investigation we observed that collaboration increased when a firm from the OpenStack ecosystem was deploying (i.e., installing or delivering) OpenStack. It occurred when Cloudscaling deployed OpenStack to KT (formerly Korea Telecom); it also happened when Mirantis helped in deploying OpenStack to AT&T, suggesting that HP, Rackspace and Canonical are forced to collaborate more with others since they are deploying OpenStack internally. Those OpenStack installations must be running smoothly, with high guaranteed up-time, enabling revenues from selling computing services to the public as Amazon, Microsoft and Google do.

# 7 Discussion, including further research

## 7.1 Homophily

The visualizations of degree-centrality Figs. 4, 5, 6, 7, 8, 9, 10 and 11 ascertained the hyper-collaborative nature of the OpenStack as its developers often collaborate with developers affiliated with competing firms. In a more positivist way, the sub-community detection process revealed a small degree of homophily in code collaboration. As visible from Fig. 13, sub-communities within the OpenStack Nova project are highly heterogeneous as they tend to include developers from many different firms. These results contradict with prior sociological studies reporting the tendency of individuals to associate and bond with similar others [34–37].

Surprisingly, the low level of homophily in code collaboration contradicts an expected social behavior. Why do not developers tend to collaborate with colleagues that should represent the same business interests within the ecosystem? What drives homophily? Is this because of the openness of FLOSS projects? Is it due to the increasing virtualization of work practices as developers collaborate with reduced face-to-face interactions? Future research should assert if the observed low level of homophily is a characteristic only of OpenStack, or more generally of FLOSS and/or other coopetitive projects.

## 7.2 Coopetition theories and business models

Our case study integrates with the established literature on coopetition. We confirm that the need for external resources is a main driving force behind the establishment of long-term cooperative relationships [121]. Additionally, through cooperation, two companies can gain access to each other's unique resources or share the cost of developing new unique resources [2, 122]. Within an open source scenario, it is an open to and networked community that fulfills the need for external resources. Moreover, according to Bengtsson and Kock, individuals within a firm can only act in accordance with one of the two logics of interaction at a time, i.e., either compete or collaborate [2]. Hence, the two logics either have to be divided between individuals within the firm or need to be controlled and regulated by an intermediate organization such as a collective association. Again, within an open source scenario, it is the project community that plays the role of such an intermediate organization. Developers must identify themselves with the project community in order to be able to collaborate with rivals in the same community. In our case, developers must identify themselves with the open source community, the OpenStack Foundation or the OpenStack community to engage in coopetitive behaviors.

Our research however finds discrepancies with prior literature on R&D coopetitive networks, which addresses alliances in a form of either joint-ventures, consortia or other arrangements where access is granted only to a few selected partners [1, 4]. In the OpenStack case, everyone is welcome to contribute to the project, and everyone is allowed to copy, sell and distribute outcomes from the project. Prior coopetition research, derived from the Nordic milk industry, proposed that coopetition activities take place far from the customer – "competitors cooperate with activities far from the customer and compete in activities close to the customer" [2]. However, as we observed code-contributions from many different end users (such as AT&T, NASA and CERN), we illustrate that in the open source arena, coopetition can also occur very close to the market and the customers.

We have assessed how revenue models affect collaboration in OpenStack: Contrary to expectations, and visible in Fig. 14 and Table 7, and with the exception of a few firms providing public cloud services (HP, Rackspace and Canonical), firms competing for the same revenue model (i.e., where rivalry is expected) tend to collaborate more than firms that do not compete for the same revenue model.

## 7.3 Understanding the dynamics of industrial FLOSS projects

We believe that the Software Engineering discipline can benefit from developing code-collaboration metrics and visualizations based on developers attributes (e.g., affiliation) as a valuable complement to established software development metrics emphasizing code size, code quality and productivity. The information that can be obtained by this means can lead to better decision making by stakeholders and investors, as well as to point out possible technical and organizational problems in the project [123].

In the case of projects with an ample number of industrial corporations, having the possibility of a transparent development process is of key importance. Nonetheless, companies are reluctant to invest significant capital and/or resources on a coopetitive project in perceived disadvantage to its competitors (i.e., fear of exploitation or other non-cooperative behaviors). Thus, this type of analysis facilitates the generation of trust and confidence among stakeholders, resulting in a win-win cooperative relationship.

Related to the above is the fact that collaborative software development processes should be *fair*. No matter how advanced the technology being developed or how involved a stakeholder might be, if some of the involved companies or individuals perceive that the project is not neutral to all stakeholders, this can arise mistrust and neglect which may end in abandonment or forking of the project. Given the proliferation of industrial FLOSS projects in recent times, future research should undertake what a *fair* development process is, incorporating process metrics that conveniently address this issue.

Also, an analysis like the one carried out enables supervisor agents, as is the case with the OpenStack Foundation in our case study, to identify possible malfunctions in the community. This can be the case of companies that do not want to be integrated into the community or to cooperate with other companies. Our idea hence expands the notion of community metrics [124, 125] from the individual developer to organizations by including affiliation information.

Practitioners from the high-tech sector can further explore the potential of using our socio-structural visualizations for better communicating the importance of their contributions to particular open source software projects. High-tech firms are already active in claiming their contributions to the open source community [126–128]; the use of the implemented visual approach can be used for communicating the social importance of a firm in a project. The same socio-structural visualizations are more valuable when complemented with rich textual descriptions of what was contributed [126] or in combination with measurable quantities of contributed source code [129].

Future research efforts from a software engineering point of view could be focused on obtaining better community cohesion metrics and the development of code collaboration metrics. They might not make sense in a small development team, but they are definitely relevant in a high-networked ecosystem where 'who works with who' is not obvious. Other ways of measuring collaboration (i.e., at the function level instead of at the file level) or other types of collaboration (i.e., reviews, bug fixing, etc.) would help understanding these type of communities better. The inclusion of weighted edges would add more tangible information on the frequency and importance of the collaboration among stakeholders.

### 7.4 Mixed method approach

Our findings suggest that by methodologically combining more ethnographic qualitative methods with social network visualizations, we can produce longitudinal and rich descriptions that enable a better understanding of competitive and collaborative issues simultaneously present and interconnected in large and complex software ecosystems. As a methodological note, and as in prior research employing similar research designs [33], we also warn that repository-driven social network measures (i.e., centrality, eigenvector centrality) or related visualizations must be interpreted carefully.

For instance from the `Grizzly` to the `Havana` release, either by summing the centrality measures of the developers affiliated to Mirantis or by interpreting Fig. 10, we could easily make wrong judgments about the importance or influence of Mirantis within the OpenStack Nova project. Thanks to complementary ethnographic knowledge we know that Mirantis has been in contractual supply relationships with AT&T, Cisco, Red Hat and NASA. Moreover, other companies contributing to the project (e. g., Intel, Ericsson, Red Hat and SAP) have equity participation within Mirantis. Hence, pure repository-driven quantitative judgments about the centrality, importance or influence of actors within a software ecosystem setting must be interpreted very carefully.

## 8 Threats to validity
### 8.1 Threats to construct validity

These threats consider the relationship between theory and observation, in case the measured variables do not provide a good measure of the actual factors. A main simplification in our modeling of reality has been how we identify collaboration in our case study. So, collaboration was narrowed to working on the same code files; other software development activities (testing, bug-fixing, translation, code-review, specification, testing, design) should be included. In addition, only modifications to the same file during a software release have accounted for collaboration, artificially limiting its scope and dimension.

Some choices regarding the SNA can also be improved. So far, we only used density as a community cohesion metric, other network-based measures for the same construct (e.g., transitivity, compactness, connectedness and distance-weighted fragmentation among others) should be considered for enhancing the rigor of this research. In addition, our method is still supported mainly by visualizing the resulting SNA graph and its attributes, although other means might be more suitable (i.e., a matrix based solution). Efforts should be also invested in exploring regressions of SNA topological properties with other measures such as activity or quality, that for the sake of parsimony have not been included in this manuscript.

### 8.2 Threats to external validity

These type of threats consider the problematic of generalizing our findings. In our research, we show evidence based on a single case study; our understanding is that it is representative of other industrial FLOSS projects, with volunteer and affiliated developers developing together a software. However, further research should replicate our method on other projects to explore other open-coopetition scenarios and to find out if our findings hold there too.

As Shihab et al. point out [130], a "frequent misconception is that empirical research within one company or one project is not good enough, provides little value for the academic community, and does not contribute to scientific development". The authors note that historical evidence shows otherwise; Flyvbjerg presents several examples from the fields of physics, economics, and the social sciences [131]. Basili et al. argue that the study of large samples or populations and of single case studies are both essential [132].

## 9 Conclusions

By employing a longitudinal design covering more than 4 years of a FLOSS ecosystem, we combined an extensive qualitative investigation, the mining of a software repository and Social Network Analysis. We addressed the question of how rival firms collaborate in OpenStack in an open source way, by analyzing the roles of firm affiliation and revenue stream. We found that management research in business ecosystems and coopetition provided powerful lenses for understanding the competitive and collaborative issues embedded in the OpenStack software evolution.

We learned that a qualitative analysis of archival data, combined with social network visualizations derived from source code repositories, provide a rich medium that enables a better understanding of software ecosystems. By addressing the initial research questions, new questions emerged regarding homophily, coopetition, user-innovation and deployment within a FLOSS ecosystems setting. By pure serendipity, we also provided a longitudinal description of how heterogeneous actors within a high-networked ecosystem (involving individuals, startups, established firms and public organizations) joint-develop an complex infrastructure for big-data in the open-source arena.

We shed some light on the potential of visualizing collaboration for supporting strategic alliance decisions in R&D projects, especially within large-scale and high-networked production scenarios. We argue that our SNA visualizations, retrieved from natural occurring source-code repositories can help stakeholders in assessing their inter-firm network positions for better decision-making regarding strategic alliances. Our methodological approach provides visualizations that support awareness of human activities in software development [133], potentially supporting decisions on how to balance cooperation and competition in a particular product or market area.

The practical importance of open-coopetition, as explored by this research, should be taken into account. Stakeholders in R&D projects have many reasons to consider coopetition in an open source fashion: The costs and risks of developing new products are divided among the cooperating companies; the time-to-market can be shorter as each company can contribute with its core competence; and 'extra' contributions can be more easily captured from third-party open source contributors (e.g., altruistic volunteers or user-contributors). All this happening under the ubiquitous pressure to innovate from collaborators, competitors and power-users.

Our results are of interest to the body of knowledge in Software Engineering, where FLOSS studies are on the research agenda. Additionally, we also integrate our research findings with management literature in business ecosystems and coopetition strategy. In order to facilitate future testing of our inductive findings from a single case, and as it was recurrently pointed that Software Engineering research also needs to engage in theory-building [134–137]. Furthermore, we advocate the need to take into account affiliation information of the developers to determine issues as the *health* of a software ecosystem or the *fairness* of its development process. Failing to identify inefficient or unfair development processes may lead to the abandonment or forking of the project by some stakeholders, may they be corporations or individuals.

Further steps are required for strength our results and consequently increase the validity of the proposed theoretical contributions. Also plenty of future research is needed for gaining further lessons from this complex case. Finally, there are also many other open-coopetition cases that remain unexplored, calling for additional research on how knowledge from business ecosystems applies to software development.

**Competing interests**
The authors declare that they have no competing interests. JGB works for Bitergia who has as customer the OpenStack Foundation. GR is co-owner of Bitergia. In order to avoid conflicts of interest, as there are professional relationships between Bitergia (2nd and 3rd author) and the OpenStack Foundation, all data collection and analysis process was executed by the first author.

**Authors' contributions**
JT carried out data-collection and analysis. All authors contributed to the writing, discussion and revising of this manuscript. All authors read and approved the final manuscript.

**Acknowledgments**
We are grateful to Daniel Izquierdo Cortázar, from Bitergia, for explaining the data of developer affiliations for the OpenStack Nova project, and keeping it up-to-date, well documented and in a research-friendly way.
Acknowledgements also for Tingting Lin, who initially motivated the mining of open-source repositories with social network analysis, Salman Mian who carefully reviewed our network visualizations. Also for Jukka Ruohonen and Marko Niemimaa with whom we had joyful discussions on how the important social concept of homophily applies to Software Engineering and Information Systems. A word also for the Professors Reima Suomi and Hannu Salmela, who supported fund-seeking initiatives that made this research possible.
The work of Jose Teixeira was partially supported by the Fundação para a Ciência e a Tecnologia (grant SFRHBD615612009), Liikesivistysrahasto (grant 3-1815) and Marcus Wallenberg Säätiö (grant open-coopetition R&D management strategy). The work of Gregorio Robles and Jesús M. González Barahona has been funded in part by the Spanish Government under project SobreSale (TIN2011- 28110) and by the Region of Madrid under project "eMadrid - Investigación y Desarrollo de tecnologías para el e-learning en la Comunidad de Madrid" (S2013/ICE-2715).
The last word is for the OpenStack community for developing a cool, open and research-friendly cloud computing infrastructure for big-data. More methodological details, data, high-resolution visualizations, and source-code are publicly available at http://users.utu.fi/joante/OpenStackSNA/.

**Author details**
[1]School of Economics (TSE), University of Turku, Turku, Finland. [2]Universidad Rey Juan Carlos, Camino del Molino s/n, 28943 Fuenlabrada, Madrid, Spain. [3]Bitergia S.L., Avenida Gregorio Peces Barba, 28918 Leganés, Madrid, Spain.